\documentclass[twocolumn, superscriptaddress, amsmath, amssymb, aps]{revtex4-1}

\usepackage{graphicx}
\usepackage{dcolumn}
\usepackage{bm}
\usepackage{xcolor}
\usepackage{mathtools}
\usepackage{physics}
\usepackage[normalem]{ulem}
\usepackage{soul}
\usepackage{braket}
\usepackage{xcolor}   

\usepackage{hyperref} 
\hypersetup{colorlinks=true, citecolor=blue}

\newcommand{\rev}[1]{{\color{black} #1}}

\begin{document}
\title{Superuniversal Statistics of Complex Time-Delays in Non-Hermitian Scattering Systems}

\author{Nadav Shaibe}
\email{Corresponding author: nshaibe@umd.edu}
 \affiliation{Maryland Quantum Materials Center, Department of Physics, University of Maryland, College Park, Maryland 20742-4111, USA}

\author{Jared M. Erb}
 \affiliation{Maryland Quantum Materials Center, Department of Physics, University of Maryland, College Park, Maryland 20742-4111, USA}
 
\author{Steven M. Anlage}
 \affiliation{Maryland Quantum Materials Center, Department of Physics, University of Maryland, College Park, Maryland 20742-4111, USA}

\date{\today}

\begin{abstract}
The Wigner-Smith time-delay of flux conserving systems is a real quantity that measures how long an excitation resides in an interaction region. The complex generalization of time-delay to non-Hermitian systems is still under development, in particular, its statistical properties in the short-wavelength limit of complex chaotic scattering systems has not been investigated. From the experimentally measured multi-port scattering ($S$)-matrices of one-dimensional graphs, a two-dimensional billiard, and a three-dimensional cavity, we calculate the complex Wigner-Smith ($\tau_{WS}$), as well as each individual reflection ($\tau_{xx}$) and transmission ($\tau_{xy}$) time-delays. The complex reflection time-delay differences ($\tau_{\delta R}$) between each port are calculated, and the transmission time-delay differences ($\tau_{\delta T}$) are introduced for systems exhibiting non-reciprocal scattering. Large time-delays are associated with \rev{scattering singularities} such as coherent perfect absorption, reflectionless scattering, slow light, and uni-directional invisibility. We demonstrate that the large-delay tails of the distributions of the real and imaginary parts of each time-delay quantity are superuniversal, independent of experimental parameters: wave propagation dimension $\mathcal{D}$, number of scattering channels $M$, Dyson symmetry class $\beta$, and uniform attenuation $\eta$. \rev{The tails determine the abundance of the singularities in generic scattering systems, and the} superuniversality is in direct contrast with the well-established time-delay statistics of unitary scattering systems, where the tail of the $\tau_{WS}$ distribution depends explicitly on the values of $M$ and $\beta$. \rev{ We relate the distribution statistics to the topological properties of the corresponding singularities. Although the results presented here are based on classical microwave experiments, they are applicable to any non-Hermitian wave-chaotic scattering system in the short-wavelength limit, such as optical or acoustic resonators.
}
\end{abstract}

\maketitle

\textit{Introduction}.---In this Letter, we consider the wave-scattering properties of \rev{generic} complex systems with a finite number of asymptotic scattering channels coupled to the outside world. The systems of interest have dimensions much larger than the wavelength of the waves, making the scattering properties extremely sensitive to details, such as boundary shapes and interior scattering centers. Such systems could be three-dimensional spaces such as rooms, two-dimensional billiards, or one-dimensional graphs, and the waves could be electromagnetic, acoustic, or quantum in nature. The complexity of wave propagation and interference is captured by the scattering $S$-matrix which transforms a vector of input excitations $\ket{\psi_{\text{in}}}$ on $M$ channels to a vector of outputs $\ket{\psi_{\text{out}}}$ as $\ket{\psi_{\text{out}}}=S\ket{\psi_{\text{in}}}$. The scattering diversity gives rise to strong dependence of the complex $S$-matrix elements as a function of excitation energy $E$. We focus in particular on the time these waves spend in the scattering region as they propagate from one scattering channel to another. An average dwell-time was introduced by Wigner \cite{Wigner1955} in the context of nuclear scattering. Smith \cite{Smith1960} later generalized this idea by inventing a lifetime matrix $Q = i\hbar S \frac{dS^\dag}{dE}$, the normalized trace of which is defined as the Wigner-Smith time-delay $\tilde{\tau}_{WS} = \frac{1}{M} \text{Tr}[Q]$, which is strictly real for unitary (flux-conserving) systems.

For a quantum mechanical wave, $\tilde{\tau}_{WS}$ can be directly related to the phase evolution of a wave packet and its group delay \cite{Zhang2011,Trabert2021}. For classical waves, the Wigner-Smith time-delay is simply the energy dependent shift in arrival time caused by the scattering interaction \cite{Asano2016}. Some examples of the physical uses of time-delay are in nuclear physics \cite{Bourgain2013,Deshmukh2021}, \rev{photoionization \cite{Fuchs2020,Rist2021,Elghazawy2023,Kheifets2023,Saalman2024}, tunneling time \cite{Jaworski1988,Winful2006}}, group delay of modes in waveguides and optical fibers \cite{Fan2005,Carpenter2015,Xiong2016,Patel2021,Mao2023}, wavefront shaping  and creation of particlelike scattering states \cite{Rotter2011,Gerardin2016,Horodynski2020,Frazier2020,Hougne2021}, radio frequency pulse propagation \cite{Smilansky2017,Smilansky2018}, radiation intensity statistics \cite{Brouwer2003,Fyodorov2023}, identifying zeros and poles of the scattering matrix \cite{Fyodorov1997bTRI,Fyodorov2017,Fyodorov2019,Osman2020,Lei2021WS,Lei2022}, characterization of disordered and biological media \cite{Genack1999,Chabanov2003,Ambichl2017,Pradhan2018,Huang2019,Hougne2021_media}, and in acoustics \cite{Patel2023}.

When studying scattering of short-wavelength excitations in complex systems, such as nuclei or irregular electromagnetic structures, it is appropriate to examine the statistical properties of time-delays, due to the sensitive energy and parametric variations of the scattering process. Random Matrix Theory (RMT) has been very successful at describing universal fluctuations of time-delays in unitary scattering systems \cite{Lehmann1995,Gopar1996,Misirpashaev1997,Fyodorov1997bTRI,Fyodorov1997Cross,Fyodorov1998,Tiggelen1999,Brouwer1999,Savin2001,Kottos2003,Mezzadri2013,Texier2013,Novaes2015,Cunden2015,Huang2020}. This body of work discovered that the probability distribution $\mathcal{P}(\tilde{\tau}_{WS})$ has certain consistent properties. One of particular interest is the shape of $\mathcal{P}(\tilde{\tau}_{WS})$ at extreme values, which correspond to modes with long scattering times. It was found independently by several groups that the distribution of the Wigner-Smith time-delay of a perfectly coupled system has an algebraic power-law decay for large values of the time-delay described by $\mathcal{P}(\tilde{\tau}_{WS}) \propto \tilde{\tau}_{WS}^{-(M\beta)/2-2}$ \cite{Seba96,Fyodorov1996,Brouwer1997,Savin2003,Ossipov2005,Kottos2005,Grabsch2018,Grabsch_2020} where $\beta$ is the Dyson symmetry class of the system \cite{Dyson1962}. A more complete review of Wigner-Smith time-delay results and analysis in unitary scattering systems can be found in Ref. \cite{Texier2016}.

Physical systems have loss/gain, and non-Hermitician effects cannot be ignored in experimental scattering data \cite{Fyodorov2004,Savin2005,Fyodorov_2005,Zheng20051Port,Zheng20052Port,Zheng2006,Hemmady2006,Hemmady2006b,Hemmady2006c,Yeh2013,Pradhan2018}. For weak attenuation, an expansion about the unitary scattering matrix is possible \cite{Beenakker2001,Fyodorov2003,Novaes2023,Novaes2023_2}, but the most general way to account for dissipation leads to a complex $\tilde{\tau}_{WS}$ \cite{Asano2016,Bohm2018,Hougne2021,Lei2021WS}. Even when the scattering matrix is unitary, the reflection and transmission submatrices are subunitary, leading to complex time-delays (CTDs) describing reflection and transmission processes \cite{Asano2016,Bredol2021}. The reflection time-delay difference between channels has been proposed as a quantity that depends only on the reflection zeros of the system \cite{Lei2022,Osman2020,Fyodorov2019}. We introduce here the complex \textit{transmission time-delay difference} in non-reciprocal scattering systems which depends solely on the system's \textit{transmission} zeros.


The presence of absorption introduces qualitatively new phenomena not seen in unitary systems, such as Coherent Perfect Absorption (CPA), in which all the energy injected into a system is completely absorbed, with no reflection or transmission on any of the scattering channels \cite{Chong2010,Fyodorov2017,Lei2020,Imani2020,Frazier2020,Hougne2021,Erb2024,Faul2024}. \rev{Also known as anti-lasing, this phenomena has drawn interest in optics \cite{Gupta2012,Pichler2019,Baranov2017}, acoustics \cite{Song2014,Meng2017}, heat transfer \cite{Li2022}, and quantum single photon systems \cite{Vetlugin2021,Vetlugin2022_Resolution,Vetlugin2022_AHOM}.} CPA has been associated with diverging Wigner-Smith time-delay \cite{Hougne2021,Huang2022}, justifying attention to the large value\rev{s of time-delay}. \rev{While $\tilde{\tau}_{WS}$ features most prominently in the literature, the other time-delays should not be forgotten, as for example} divergences of the reflection time-delay correspond to reflectionless scattering modes (RSM) \cite{Sol2023,Jiang2024,Faul2024}, and the divergences of the transmission time-delay difference could identify points of uni-directional invisibility \cite{Lin2011,Kurter2011,Peng2014}. \rev{In Fig.~\ref{combschem}, we show the real parts of the (b) Wigner Smith, (c) Reflection, and (d) Transmission time delay from an experimental microwave graph (discussed below) over a range of frequency and phase shift along one bond. The purple symbols mark the divergences of the time-delays and their corresponding singularities. When looking at a probability distribution of delay from a statistical ensemble, the tail of the distribution gives a measure of the abundance of these extreme phenomena, which show rich potential for applications.}


Recently, Chen, \textit{et al.} showed that \rev{$\mathcal{P}($Re$[\tilde{\tau}_{WS}])$ and $\mathcal{P}($Im$[\tilde{\tau}_{WS}])$} of perfectly coupled $M=2$ channel, non-reciprocal ($\beta=2$) \rev{graphs ($\mathcal{D}=1$)} with finite uniform absorption strength $\eta$ \rev{have algebriac} $-3$ power-law tails. They predicted that this should be true \rev{of $\mathcal{P}(\tilde{\tau}_{WS})$} for \rev{any} perfectly coupled non-Hermitian system \cite{Lei2021Stats}. In this Letter, we examine the statistics of each CTD with a focus on the asymptotic behavior of their distributions. We \rev{consider} four \rev{global} parameters which could dictate the distribution tails: (i) dimension for wave propagation $\mathcal{D}$, (ii) number of channels $M$, (iii) Dyson class $\beta$, and (iv) uniform dissipation $\eta$.

\begin{figure}[h]
\includegraphics[width=0.48\textwidth]{fig1_combined_surface_schematic.png}
\caption{\rev{a)} Schematic of an experimental $M=2$ tetrahedral graph ($\mathcal{D}=1$). Four of the six bonds of the graph are comprised of phase shifters. Each node can be an SMA T-adapter/junction or microwave circulator. \rev{Two-parameter heatmap of (b) $\text{Re}[\tilde{\tau}_{WS}]$, (c) $\text{Re}[\tilde{\tau}_{xx}]$, and (d) $\text{Re}[\tilde{\tau}_{xy}]$ from a reciprocal ($\beta=1$) tetrahedral microwave graph with $\Delta=46$ MHz and $\tau_H=21$ ns. Purple symbols indicate locations of diverging CTD at (b) CPAs, (c) zero reflection, (d) and zero transmission points.}}
\label{combschem}
\end{figure}

We first provide the theory and definitions of CTD in non-Hermitian systems, then present the systems experimentally investigated, and finally we discuss the superuniversal $-3$ power-law tail of the CTD probability distributions. 
 

\textit{Theory}.---In the Heidelberg approach to wave scattering \cite{VERBAARSCHOT1998,SOKOLOV1989,Fyodorov1997bTRI,FSav11,Kuhl2013,Schomerus2015}, the non-Hermitian effective Hamiltonian of a scattering environment $\mathcal{H}_{\text{eff}}= H-i\Gamma_W$
is constructed from the $N\cross N$ Hamiltonian $H$ describing the closed system and the $N\cross M$ matrix $W$ of coupling coefficients between the $N$ modes of $H$ and the $M$ scattering channels, where $\Gamma_W = \pi W W^\dag$. The energy-dependent scattering matrix $S(E)$ takes the form
\cite{SOKOLOV1989,Fyodorov1997bTRI,Kuhl2013}
\begin{equation*}
    S(E) = 1_{M\cross M}-2\pi i W^\dag\frac{1}{E-\mathcal{H}_\text{eff}}W = \begin{pmatrix}
S_{11} & \dots & S_{1M}\\
\vdots & \ddots & \vdots \\
S_{M1} & \dots  & S_{MM}
\end{pmatrix}.
\end{equation*}
One can account for a spatially uniform attenuation of the waves with rate $\tilde\eta$ by making the substitution $E\rightarrow E+i\tilde\eta$ and evaluating the resulting subunitary $S$-matrix at complex energies. Experimentally, such a subunitary scattering matrix for an $M$-channel system is measured as $S(E) $ in terms of $M^2$ energy-dependent complex reflection ($S_{xx}$) and transmission ($S_{xy}$) submatrices where $x\neq y$ ($x,y \in [1,\dots, M]$) \cite{Kang2021,Huang2022,Fisher1981,Rotter2017}. \rev{It should be noted here that $\tilde{\eta}$ is the uniformly imposed mode bandwidth, and} describing uniform loss in this manner is phenomenological, making no special assumptions about the $S$-matrix \cite{Buttiker1986,Brouwer1996,Kuhl2013}.

The Wigner-Smith time-delay extended to non-Hermitian systems has a natural definition as a complex quantity \cite{Lei2021WS}:
\begin{equation}
    \tilde{\tau}_{WS} := \frac{-i}{M}\frac{\partial}{\partial E}\text{log}[\text{det}S(E+i\tilde{\eta})].
    \label{DefWS}
\end{equation}
$\tilde{\tau}_{WS}$ can also be written as a sum of Lorentzian functions of energy in terms of the $S$-matrix poles and zeros, which are the complex eigenvalues of $\mathcal{H}_{\text{eff}}$ and $\mathcal{H}_{\text{eff}}^\dag$ respectively \cite{Lei2022,Ma2023}.

While $\tilde{\tau}_{WS}$ takes into account the entire scattering matrix, it is also important to consider the time-delays of individual reflection and transmission processes \cite{Sebbah1999,Bemmel2001,Asano2016,Bohm2018,Bredol2021,Bialous2021,Genack2024,Lei2024}. The CTDs for reflection at channel $x$ and transmission to channel $x$ from channel $y$ are given by \cite{Lei2022}
\begin{subequations}
\begin{align}
          \tilde{\tau}_{xx} := -i\frac{\partial}{\partial E}\text{log}[S_{xx}] \label{dR}\\
    \tilde{\tau}_{xy} := -i\frac{\partial}{\partial E}\text{log}[S_{xy}] \label{dT}
\end{align}
\end{subequations}
 in direct analogy with $\tilde{\tau}_{WS}$, and can also be written as sums of Lorentzian functions of energy \cite{Yuhao2021,Lei2022}.

Since the Lorentzian terms arising from the poles are identical for each reflection and transmission time-delay, we can define new quantities, the reflection and transmission time-delay \textit{differences}, which are expected to be independent of the S-matrix poles \cite{Lei2022,Osman2020,Fyodorov2019}:
\begin{subequations}
\begin{align}
    \tilde{\tau}_{\delta R} := \tilde{\tau}_{xx} - \tilde{\tau}_{yy} =-i\frac{\partial}{\partial E}\text{log}[\frac{S_{xx}}{S_{yy}}]    \label{DefdR}\\
    \tilde{\tau}_{\delta T} := \tilde{\tau}_{xy} - \tilde{\tau}_{yx} =-i\frac{\partial}{\partial E}\text{log}[\frac{S_{xy}}{S_{yx}}].\label{DefdT}
\end{align}
\end{subequations}
In an arbitrary system, usually $S_{xx}\neq S_{yy}$ so $\tilde{\tau}_{\delta R}$ is almost always non-zero. In contrast, to have a non-trivial $\tilde{\tau}_{\delta T}$, the system must have non-reciprocal transmission. This can be accomplished, for example, with electromagnetic waves propagating through magnetized ferrite materials \cite{So1995,Stoffregen1995,WuEigen98,Chung00,Schanze2005,Lawniczak2009,Lawniczak2010,Lawniczak2011,Lawniczak2019,Zhang2023}. The transmission time-delay difference written as sums of Lorenztian functions of energy is given for the first time in Eqs.~\ref{realtdiff}-\ref{imagtdiff}.

In highly over-moded structures, CTDs fluctuate strongly in energy (or equivalently frequency $f$), with both the real and imaginary parts taking on positive and negative values \cite{Asano2016,Lei2021WS,Hougne2021,Lei2022} (see Fig.~\ref{sampling}(a) for a representative example). Since CTD is very sensitive to perturbations of an over-moded scattering system, we examine the distribution of time-delays from a statistical ensemble of many similar systems \rev{to make general statements}.


\textit{Experiment}.---Microwave experiments have proven to be ideal platforms for investigating fluctuations in wave scattering properties of complex systems \cite{Doron1990,Stockmann1990,StockBook99, Richter2001,Hul2012,Kuhl2013,Gradoni2014,Dietz2015}. The experimental data presented in this paper comes from $M\times M$ scattering matrices $S$ collected through the use of calibrated Keysight PNA-X N5242A and PNA-X N5242B microwave vector network analyzers. The measured scattering parameters from the PNA contain information about the coupling of the system, as well as direct processes (which include short orbits between ports). These non-universal effects are removed \rev{from an ensemble} through application of the random coupling model (RCM) normalization process, which results in an $S$-matrix with perfect coupling, and reduces the effects of short orbits on the statistics \cite{Hemmady2005,Zheng20051Port,Zheng20052Port,Zheng2006,Hart2009,Yeh2010, Lawniczak2012,Hemmady2012,Lee2013,Yeh2013,Gradoni2014,Auregan16,Dietz2017,Fu2017}.\rev{~To make a statistical ensemble, tunable perturbers that can change local conditions between realizations are embedded in the scattering systems.  High quality determinations of $\eta$ are extracted from the statistical ensembles; details can be found in Supp. Mat. \cite{SuppMatt} Sec.~\ref{SupMat_Systems}C.}

Because physical systems have system-specific shape and size, the value in seconds of any time-delay $\tilde{\tau}$ is dependent to some extent on irrelevant details. To isolate the universally-fluctuating properties, it is necessary to normalize $\tilde{\tau}$ by the Heisenberg time $\tau_{H} = \frac{2\pi}{\Delta}$, where $\Delta$ is the mean mode spacing of the closed system \cite{Lei2021Stats,Lyuboshitz1977,Pierrat2014}. The scaled time-delays $\tau:=\tilde{\tau}/\tau_H$ discussed in this Letter are dimensionless quantities independent of the specific system measured. We also normalize the absorption rate of the system to a dimensionless quantity in units of the Heisenberg time $\eta := \tau_H\tilde{\eta}$.

Our $\mathcal{D}=1$ system is an irregular tetrahedral graph, a well-studied graph topology \cite{Kottos1997,Hul2004,Li2017,Lei2020,Hofmann2021,Lu2023,Tornike2023,Dietz2024}, with \rev{tunable} phase shifters on the bonds \rev{to manipulate the wave interference conditions} \cite{Rehemanjiang2018,Dietz2017,Che2021,Che2022,Lawniczak2023}, depicted schematically in Fig.~\ref{combschem}\rev{(a)}. For $\mathcal{D}=2$ and $\mathcal{D}=3$ we used a two-dimensional ray-chaotic 1/4-bow-tie billiard \cite{Stockmann1990,Stoffregen1995,Alt1998,Dietz2006,Dietz2015} and a three-dimensional ray-chaotic cavity \cite{Deus1995,Alt97,Tait2011,Xiao2018,Hougne2020,Gros2020,Frazier2020,Frazier2022}, each containing voltage controlled\rev{, varactor-loaded} metasurfaces \rev{which can significantly alter the boundary conditions of the systems.} \cite{Chen2016,Elsawy2023,Frazier2022,Sleasman2023,Erb2024}. A schematic of the billiard used in the experiment is shown in Fig.~1 of Ref. \cite{Erb2024}, and a photograph of the cavity and metasurface is given in Fig.~\ref{PhotoGigabox.}. The systems were measured with $M=1,2,3$ scattering channels. The Dyson symmetry class $\beta$ of a system can be changed from $1$ to $2$ through inclusion of a magnetized ferrite in the microwave propagation path, such as including a circulator at one of the nodes in a graph \cite{Dietz2007,Dietz2009,Lawniczak2009,Lawniczak2010,Lawniczak2011,Bittner2014,Castaneda2022}. We also consider a graph that has a circulator on every node connected to a scattering channel, but nowhere else (see Fig.~\ref{SchematicNRC}), which we term ``non-reciprocal coupling" (NRC), as a special case of $\beta = 2$. Each system has its own intrinsic uniform absorption strength $\eta$ which is frequency dependent \cite{Lei2022}, therefore by measuring in different frequency bands we can systematically vary the value of $\eta$. The uniform attenuation can also be increased by uniformly distributing absorbers in a cavity \cite{Hemmady2005,Hemmady2005b,Hemmady2006,Hemmady2006c} or attenuators on the bonds of a graph \rev{\cite{Hul2005,Lawniczak2023,Lawniczak2019}}. The absorption rate of the systems considered here varied from $\eta=1.8$ to $\eta=50$, creating data in \rev{what are considered} the low, moderate, and high absorption strength regimes \cite{Sanchez2003,Hul2004,Hemmady2005,Fyodorov2004,Savin2005,Lawniczak2008,Lawniczak2019}.


More details on the specific experimental systems as well as the creation of statistical ensembles can be found in Supp. Mat. \cite{SuppMatt} Sec.~\ref{SupMat_Systems}.


\textit{Discussion}.---We present the probability distribution functions (PDFs) of the CTDs calculated using Eqs (\ref{DefWS}-\ref{DefdT}) from the measured ensembles of $S$-matrix data that have been RCM-normalized to establish perfect coupling at all frequencies.

\begin{figure}[th]
\includegraphics[width=0.48\textwidth]{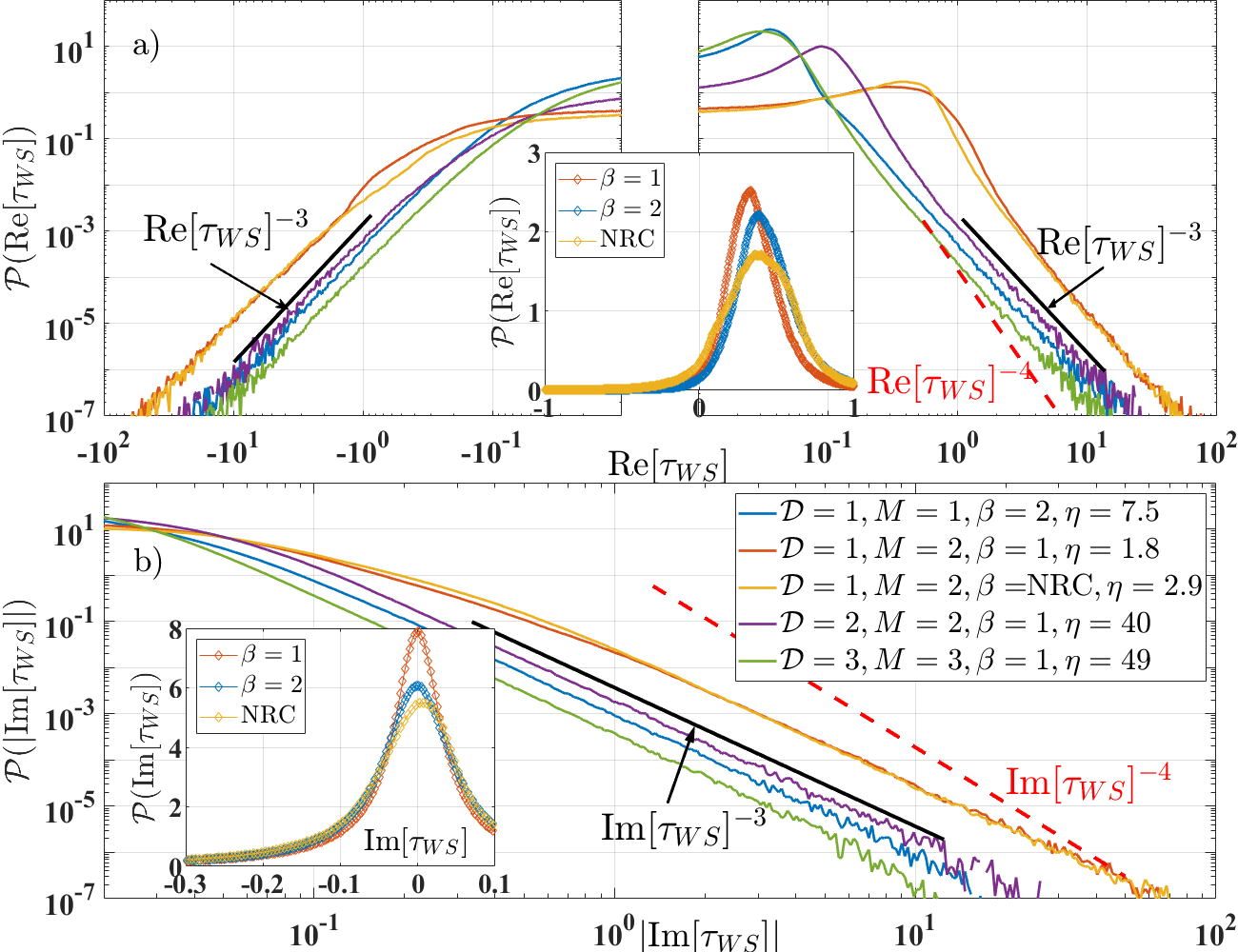}
\caption{PDFs of (a) Re$[\tau_{WS}]$  and (b) $\lvert$Im$[\tau_{WS}]$$\rvert$ Wigner-Smith CTD from ensembles of experimental scattering systems with different values for the four parameters $\mathcal{D},M,\beta,\eta$. Black reference line characterizes asymptotic behavior as $\tau_{WS}^{-3}$ power-law for all distributions.  Dashed red line depicts $-4$ power-law showing deviation from distributions. Insets depict representative distributions of a $\beta=1, \beta=2$ and NRC system on a linear scale, all with $\mathcal{D}=1,M=2$ and approximately equal $\eta$.}
\label{WS}
\end{figure}

Figure \ref{WS}(a) shows the distributions of Re$[\tau_{WS}]$ for five systems with different values for the four experimental parameters, $\mathcal{D},M,\beta,\eta$, and Fig. \ref{WS}(b) shows the PDFs of $|$Im$[\tau_{WS}]|$. The black lines are not fits, but $-3$ power-laws placed nearby to characterize the large-delay tail behavior. For large values of $\tau_{WS}$, all distributions have parallel tails with the same slope as the black reference line. 
The dashed red line in Fig.~\ref{WS} depicts a $-4$ power-law, \rev{which a Hermitian $\beta=2,M=2$ system would have}, but clearly deviates from the data.


Individual complex reflection and transmission time-delay distributions calculated from Eqs.~\ref{dR}-\ref{dT} are presented in Supplemental Material Sec.~\ref{SupMat_Individual}. The tails of these distributions, both real and imaginary, have a $-3$ power-law. We also look at the distributions of the CTD differences, as defined in Eqs.~\ref{DefdR}-\ref{DefdT}. Fig.~\ref{deltaR}  \rev{in the Supp. Mat.} shows $\mathcal{P}(\tau_{\delta R})$ of five systems with varying parameters. 
Fig.~\ref{deltaT} shows the distributions of \rev{the new quantity} $\tau_{\delta T}$ for six systems with $\mathcal{D}=1$ and $\beta=2$. For this work all measured broken-reciprocity systems were graphs \rev{($\mathcal{D}=1$). More detail on $\tau_{\delta T}$ is available in Supp. Mat. Sec.~\ref{SupMat_deltaT}.}

\begin{figure}[h]
\includegraphics[width=0.48\textwidth]{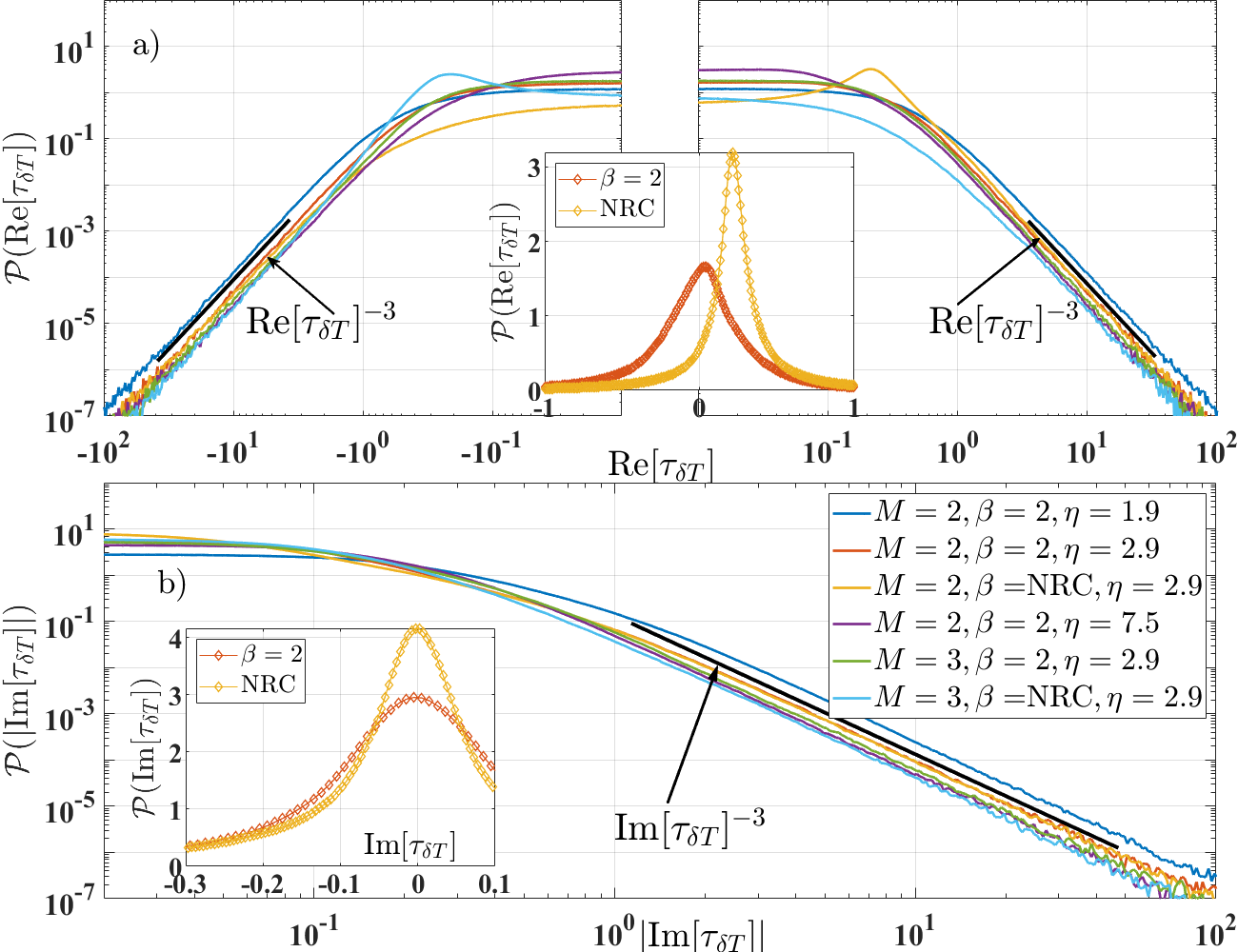}
\caption{PDFs of (a) Re$[\tau_{\delta T}]$ and (b) $\lvert$Im$[\tau_{\delta T}]$$\rvert$ transmission time-delay difference from ensembles of experimental scattering systems with different values for $M$ and $\eta$, and two different non-reciprocal ($\beta=2$, NRC) graphs. All distributions come from graphs ($\mathcal{D}=1$). Insets and reference lines serve the same purpose as in Fig.~\ref{WS}. }
\label{deltaT}
\end{figure}

\begin{table}
    \centering
    \begin{tabular}{|c|c|c|c|}
    \hline
        \rev{CTD $\tau$} & \rev{\# Distributions Fit }&\rev{$\alpha_{\text{Im}}$} &\rev{$\alpha_{\text{Re}}$} \\
        \hline
        \rev{$\tau_{WS}$ }& \rev{27} & \rev{$2.99\pm0.03$} &\rev{ $2.984\pm0.006$} \\
        \hline
        \rev{$\tau_{xx}$} &\rev{ 49} & \rev{$3.00\pm0.02$} & \rev{$2.982\pm0.006$} \\
        \hline
        \rev{$\tau_{xy}$ }&\rev{ 64} & \rev{$3.00\pm0.014$} & \rev{$3.010\pm0.008$} \\
        \hline
        \rev{$\tau_{\delta R}$} & \rev{32} & \rev{$3.030\pm0.007$} & \rev{$2.994\pm0.008$} \\
        \hline
        \rev{$\tau_{\delta T}$} & \rev{44} & \rev{$3.011\pm0.009$} & \rev{$3.006\pm0.005$}\\
        \hline
    \end{tabular}
    \caption{\rev{Table of power-law fits to experimental time-delay distributions. First and second columns are the kind of CTD and the number of distributions fit to the form $\mathcal{P}(_{\text{Im}}^{\text{Re}}[\tau]) = ~_{\text{Im}}^{\text{Re}}[\tau]^{-\alpha}$ over the range $10^{-6}\leq\mathcal{P}(_{\text{Im}}^{\text{Re}}[\tau])\leq10^{-3}$. The third (fourth) column is the mean fit for tails of the distributions of the real (imaginary) components with uncertainty.}}
    \label{TailTable}
\end{table}

Across Figs.~\ref{WS}-\ref{deltaT}, and Figs.~\ref{Rxx}-\ref{deltaR}, we show empirical evidence for a superuniversal $-3$ power-law tail behavior \rev{(see Table \ref{TailTable} for numerical values)} for every CTD ($\tau_{WS},\tau_{xx},\tau_{xy},\tau_{\delta R},\tau_{\delta T}$) distribution of perfectly coupled non-Hermitian systems. The tail on all these distributions implies that no CTD quantity has a finite variance. Compare this to the established variance of $\tau_{WS}$ for Hermitian systems which diverges only for $M\beta=2$ \cite{Mezzadri2013,Kuipers2014,Grabsch2018}.

\rev{According to these results, extreme time-delay phenomena and their associated scattering singularities (such as CPAs, RSMs, etc.) are not events that require careful engineering. A complex scattering system with tunable local parameters, such as metamaterials \cite{Erb2024,Chen2016,Frazier2020,Frazier2022,Elsawy2023,Sleasman2023,Faul2024}, should host many such singularities, as was seen to be the case for $S$-matrix exceptional point degeneracies in \cite{Erb2024_TopSing}.} Such controlled perturbations can be accomplished not only with \rev{microwave photonics}, but also in acoustics \cite{Stein2022,Huang2024,Cummer2016} and optics \cite{Chong2010,Wan2011,Pichler2019}.

\rev{While all the results discussed in this paper are for electromagnetic waves, the results apply generally to all classical wave scattering systems. It has also been noted that there is a direct analogy between quantum weak measurements and CTD  \cite{Aharonov1988,Asano2016,Solli2004,Dennis2012}, suggesting that the statistical results presented here extend also to the realm of quantum measurements.} 

In Supplemental Material Sec. \ref{SupMat_RMT} we demonstrate that the $-3$ power-law tail is also displayed in the distributions of CTD calculated from RMT numerical data, using the same set of parameters used in the experiment. The statistics of time-delays derived from $S$-matrix data with \rev{(imperfect)} frequency dependent coupling are shown in Supplemental Materials \cite{SuppMatt} Section \ref{SupMat_Coup}.


Concerning theory, \rev{the $-3$ power-law tails were predicted for $\mathcal{P}(\text{Re}[\tau_{WS}])$, $\mathcal{P}(\text{Im}[\tau_{WS}])$ \cite{Lei2021Stats}, and $\mathcal{P}(\text{Re}[\tau_{\delta R}])$ \cite{Fyodorov2019,Osman2020} through considerations of the $S$-matrix poles and zeros for the first two, and reflection zeros for the latter. We detail non-rigorous, plausibility extensions of these predictions to $\mathcal{P}(\tau_{xx})$ and $\mathcal{P}(\tau_{\delta T})$ in Supp. Mat. Sec.~\ref{SupMat_Theory}. However, there is no way known to us to do something similar for $\mathcal{P}(\tau_{xy})$ \cite{Huang2022,Huang2022_note}. Since there is a -3 power-law tail observed for all time-delay quantities, it can be inferred that there should be an underlying reason, independent of the various poles and zeros of the scattering matrix.}


\rev{We suggest a possible explanation for the observed, unpredicted superuniversality. The tails of the time-delay distributions essentially describe the relative likelihood of time-delay divergences, which occur at scattering singularities. These singularities are zeros of complex scalar functions, which are topologically stable \cite{Berry2000}, meaning they do not appear or disappear spontaneously, but only through pairwise-creation or -annihilation with a partner of opposite winding number \cite{Erb2024_TopSing,Neu90}. In a 2-parameter space, see Fig.~\ref{combschem}(b-d), the time-delay divergences occur at singular points. We propose that the $-3$ power-law tail is generic to the distributions of any such quantity with these topological defects, including CTD.
Another example is the quantum weak measurement value calculated by Solli \textit{et al} \cite{Solli2004}  which diverges at topologically stable singularities of the response function.
}



\textit{Conclusion}.---In this Letter, we have experimentally demonstrated that the ensemble distribution of every CTD quantity ($\tau_{WS},\tau_{xx},\tau_{xy},\tau_{\delta R},\tau_{\delta T}$) of non-Hermitian complex scattering systems have simple asymptotic behavior with a $-3$ power-law tail, and that this feature is superuniversal regardless of (i) wave propagation dimension $\mathcal{D}$, (ii) number of scattering channels $M$, (iii) Dyson symmetry class $\beta$, and (iv) uniform absorption strength $\eta$, at least in the range that we have studied experimentally and through RMT numerics. This result is unexpected on the basis of theory for Hermitian scattering systems. Further, we have introduced the transmission time-delay difference $\tau_{\delta T}$, appropriate for non-reciprocal systems. The simple \rev{asymptotic} form the distributions take implies an abundance of singular events in arbitrary non-Hermitian scattering systems, such as acoustic/optical/\rev{photonic} resonators. \rev{We suggest that the origin of the observed superuniversality is tied to the topological properties of the singularities associated with the extreme ends of the distribution.}


\bigskip
\textbf{Acknowledgements} We acknowledge Dr.~Lei Chen for foundational work and Prof.~Yan Fyodorov for insightful discussions on complex time-delay. This work was supported by NSF/RINGS under grant No. ECCS-2148318, ONR under grant N000142312507, and DARPA WARDEN under grant HR00112120021. 


\bigskip







\pagebreak

\setcounter{figure}{0}
\setcounter{equation}{0}
\setcounter{section}{0}
\makeatletter
\renewcommand{\figurename}{Fig.}
\renewcommand{\thefigure}{S\arabic{figure}}   

\renewcommand{\tablename}{Table}
\renewcommand{\thetable}{S\Roman{table}}    

\renewcommand{\theequation}{S\arabic{equation}}    

\renewcommand{\thesection}{S\arabic{section}}
\makeatother

\title{Superuniversal Statistics of Complex Time-Delays in Non-Hermitian Scattering Systems}

\maketitle

\date{\today}


\begin{center}

SUPPLEMENTARY MATERIAL 

\vspace{0.4cm}



Superuniversal Statistics of Complex Time-Delays in Non-Hermitian Scattering Systems

\medskip
\vspace{0.4cm}
Nadav Shaibe, Jared M. Erb, and Steven M. Anlage

\end{center}
\vspace{1cm}

Here we provide the reader with additional details relevant to the superuniversal complex time-delay statistics. Section \ref{SupMat_Systems} has an explanation of the experimental systems and the methods used to create ensemble statistics. Section \ref{SupMat_deltaT} has a more explicit discussion of the transmission time-delay difference. Section \ref{SupMat_Individual} has the experimental distributions of the individual complex reflection and transmission time-delays\rev{, as well as the distributions of the reflection time-delay difference}. In Section \ref{SupMat_RMT}, we discuss the use of Random Matrix Theory numerics to examine generic CTD statistics, and the long-tail behavior. In Section \ref{SupMat_Lim}, we discuss two avenues we explored to identify the limits of the superuniversal time-delay statistics. \rev{Finally, in Section \ref{SupMat_Theory}, we provide a short extension of the existing theory for $\mathcal{P}(\tau_{WS})$ and $\mathcal{P}(\tau_{\delta R})$ to the other complex time-delay distributions.}

\section{Experimental Systems} \label{SupMat_Systems}

In this section, we provide details on the systems used in the experiment and the creation of statistical ensembles. We also discuss the limitations of the experiment, and point out possible pitfalls for others who might pursue similar experiments.

\subsection{One-Dimensional Graph}
For $\mathcal{D}=1$, we measured a series of irregular microwave graph ensembles with $M=1,2,3$ ports connecting the system to the outside world. Microwave graphs experiments \cite{Hul2004,Lawniczak2008,Hul2012,Lei2020,Hofmann2021,Castaneda2022,Lu2023} have a number of advantages over other scattering systems. The mean mode spacing $\Delta = \frac{c}{2L_e}$ is energy independent and can be varied by changing the overall length $L_e$ of the network. \rev{For this work, the graphs used had $L_e$ values which led to Heisenberg times $\tau_H$ on the order of $10$ ns}. One can also precisely vary the uniform absorption by measuring in different frequency bands, and by inclusion of microwave attenuators in the bonds of the graph. Both reciprocal ($\beta=1$) and non-reciprocal ($\beta=2$) regimes are accessible through the use of magnetized ferrite circulators \cite{Dietz2007,Dietz2009,Lawniczak2009,Lawniczak2010,Lawniczak2011,Bittner2014} which break reciprocity for microwave flow. Finally, graphs are remarkably stable and resistant to environmental perturbations allowing for large-scale data collection over a long period of time without the parameters of the system drifting from nominal values. Graphs have disadvantages for statistical studies, which higher dimensional systems such as billiards do not share, including significant reflection at the nodes, which causes trapped modes on bonds \cite{Fu2017,Tornike2023}, and persistent short orbits \cite{Lawniczak2012,Dietz2017}. These effects can cause statistical results from graphs to deviate from Random Matrix Theory (RMT) predictions \cite{Tornike2023,Ma2023}\rev{, such as the small differences seen between the numerical and experimental distributions of normalized impedance in Fig.~\ref{RCMStats}}.

\begin{figure}[ht]
\includegraphics[width=0.48\textwidth]{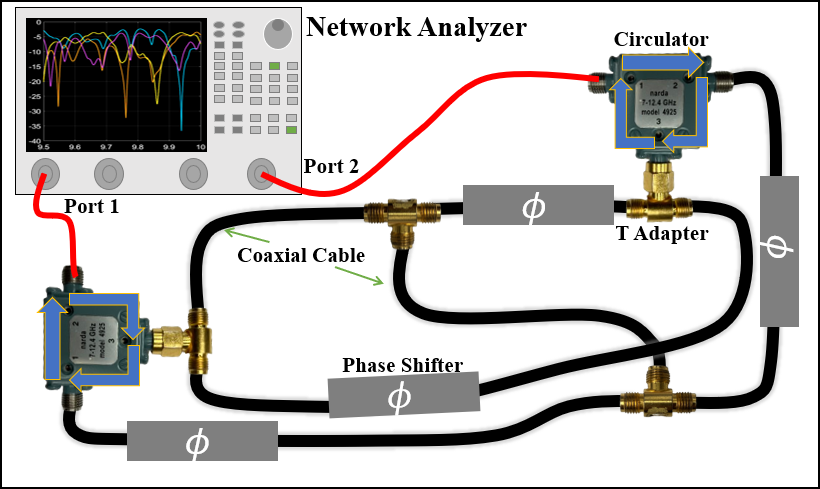}
\caption{Schematic of an experimentally realized irregular $M=2$ tetrahedral graph with two channels and non-reciprocal coupling ``NRC". Every node connected to a scattering channel is a circulator, and there are no circulators on internal nodes.}
\label{SchematicNRC}
\end{figure}

The microwave graphs used in this experiment are constructed with coaxial cables and microwave phase shifters connected to SMA T-type adapters in the topology of a tetrahedron, as shown in Fig.~\ref{combschem}\rev{(a)}. This is a standard graph employed in the wave chaos literature \cite{Kottos1997,Hul2004,Li2017,Lei2020,Hofmann2021}. The T-adapters act as the $\nu=4$ nodes of the graph and are the only places where the waves interact. $M$ of the nodes are connected to a Network Analyzer for measurement of the scattering matrix. An ensemble is created through adjusting the setting of four NARDA-Miteq P1507D-SM24 phase shifters, which make up parts of the bonds of the graph, and can be digitally controlled simulatenously through the use of a Trinamic TMCM-6110 PCB. The PCB is connected to the measurement computer that writes the commands, and to a Keithley DC power supply to provide current to the phase shifter stepper motors. Changing the lengths of individual bonds of the graph under the constraints that all six lengths be incommensurate, and that the overall length $L_e$ (and hence $\Delta$ \rev{and $\eta$}) remain fixed between realizations, results in a high quality statistical ensemble \cite{Dietz2017,Che2021,Che2022,Lawniczak2023}. \rev{This way of generating graph ensembles is better than physically replacing cables, as was done in Refs.~\cite{Lei2021Stats} and \cite{Fu2017}, because: (i) it allows for a larger ensemble with more realizations for better statistics, and (ii) it preserves the global properties of the graphs so different realizations are more comparable.}

The number of scattering channels $M$, the amount of absorption $\eta$, and the presence/absence of reciprocity $\beta$ differentiate the ensembles, which are all made of the same set of $715$ unique phase shifter settings. We also consider a special case of non-reciprocity which we refer to as the ``non-reciprocal coupling'' (NRC) case, where every port node is a circulator, shown explicitly in Fig.~\ref{SchematicNRC}. This means waves can only enter and leave the graph through particular paths, and leads to different statistics compared to a graph with circulators only on internal (non-port) nodes of the graph.

\subsection{Higher Dimensional Systems}

\begin{figure}[ht]
\includegraphics[width=0.48\textwidth]{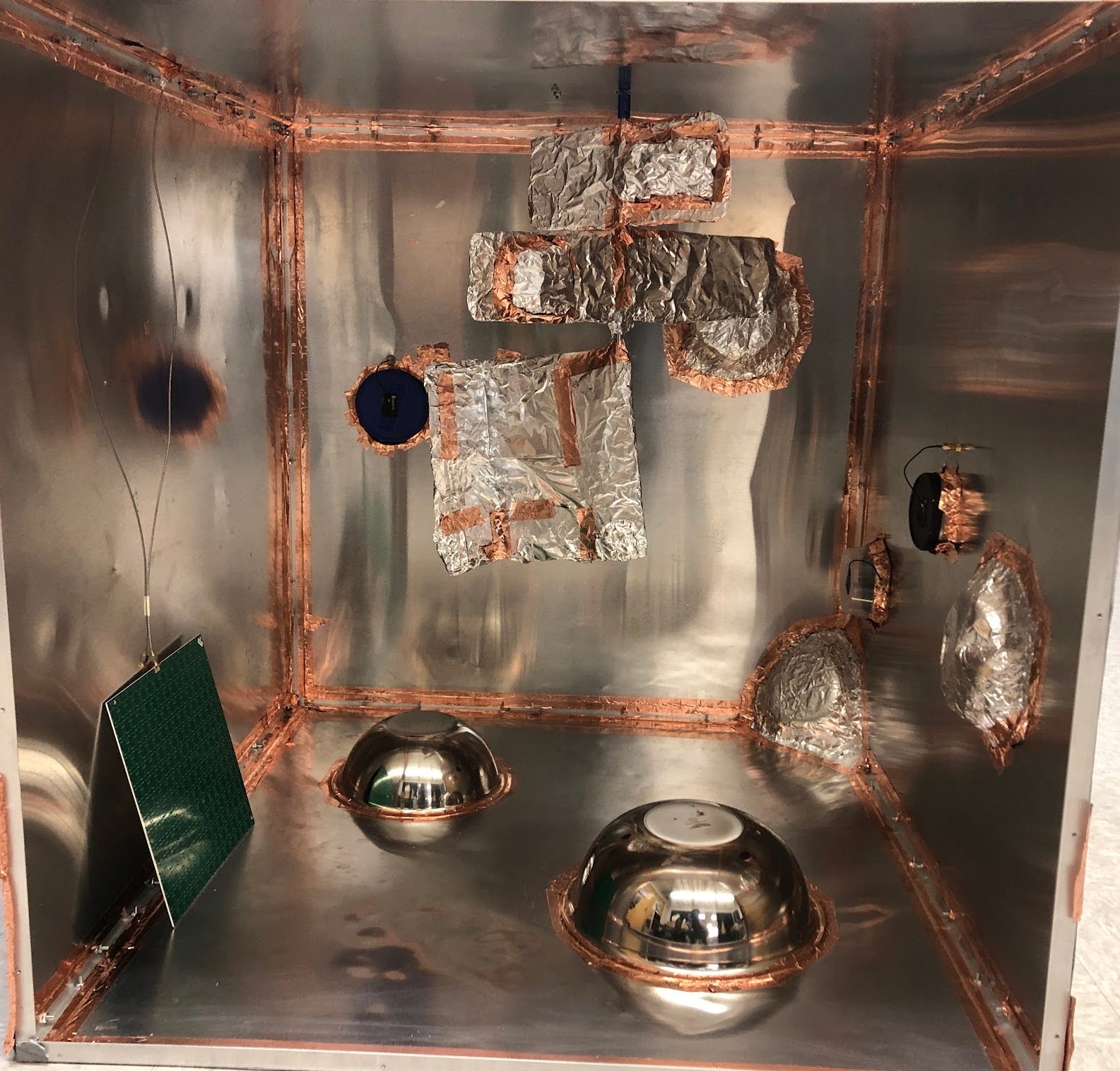}
\caption{Photograph of ray-chaotic $\mathcal{D}=3$ dimensional cavity with a tuneable metasurface (green rectangle leaning against the left wall). Line of sight between each port is broken by the use of sheets of Aluminum foil or metal bowls, which also increase the complexity of ray trajectories.}
\label{PhotoGigabox.}
\end{figure}

In addition to the graphs, we also measured the scattering matrix of $\mathcal{D}=2$ a 1/4-bow-tie billiard \cite{Stockmann1990,Alt1998,Dietz2006,Dietz2015,Xiao2018} and $\mathcal{D}=3$ a ray-chaotic cavity (see Fig.~\ref{PhotoGigabox.})\cite{Deus1995,Tait2011,Hougne2020,Gros2020,Frazier2020,Frazier2022}. An ensemble is created through adjustment of the voltage applied to varactor diode metasurfaces attached to the walls of the billiard/cavity. The metasurfaces were fabricated by the Johns Hopkins University Applied Physics Laboratory, and vary the reflection phase and amplitude of incident waves, thus altering the boundary conditions \cite{Chen2016,Elsawy2023,Sleasman2023,Erb2024}. 

The mean mode spacing $\Delta$ has a more complicated form for higher-dimensional systems, and is no longer frequency independent as in the case of one-dimensional systems. For a two-dimensional billiard, the mean mode spacing can be written as $\Delta = \frac{c}{kA}$ in terms of area $A$ and wavenumber $k$, and for a three-dimensional cavity the mean mode spacing is $\Delta = \frac{\pi c}{2k^2V}$, where $V$ is the volume of the scattering system. This complication is off-set by the fact that two and three-dimensional ray-chaotic cavities have statistics that are well described by RMT, unlike graphs. \rev{The billiard and cavity measured in this work had Heisenberg times $\tau_H$ on the order of $100$ ns and $1000$ ns, respectively.}

\subsection{Determination of the Degree of Absorption for an Ensemble} 

\begin{figure}[bht]
\includegraphics[width=0.48\textwidth]{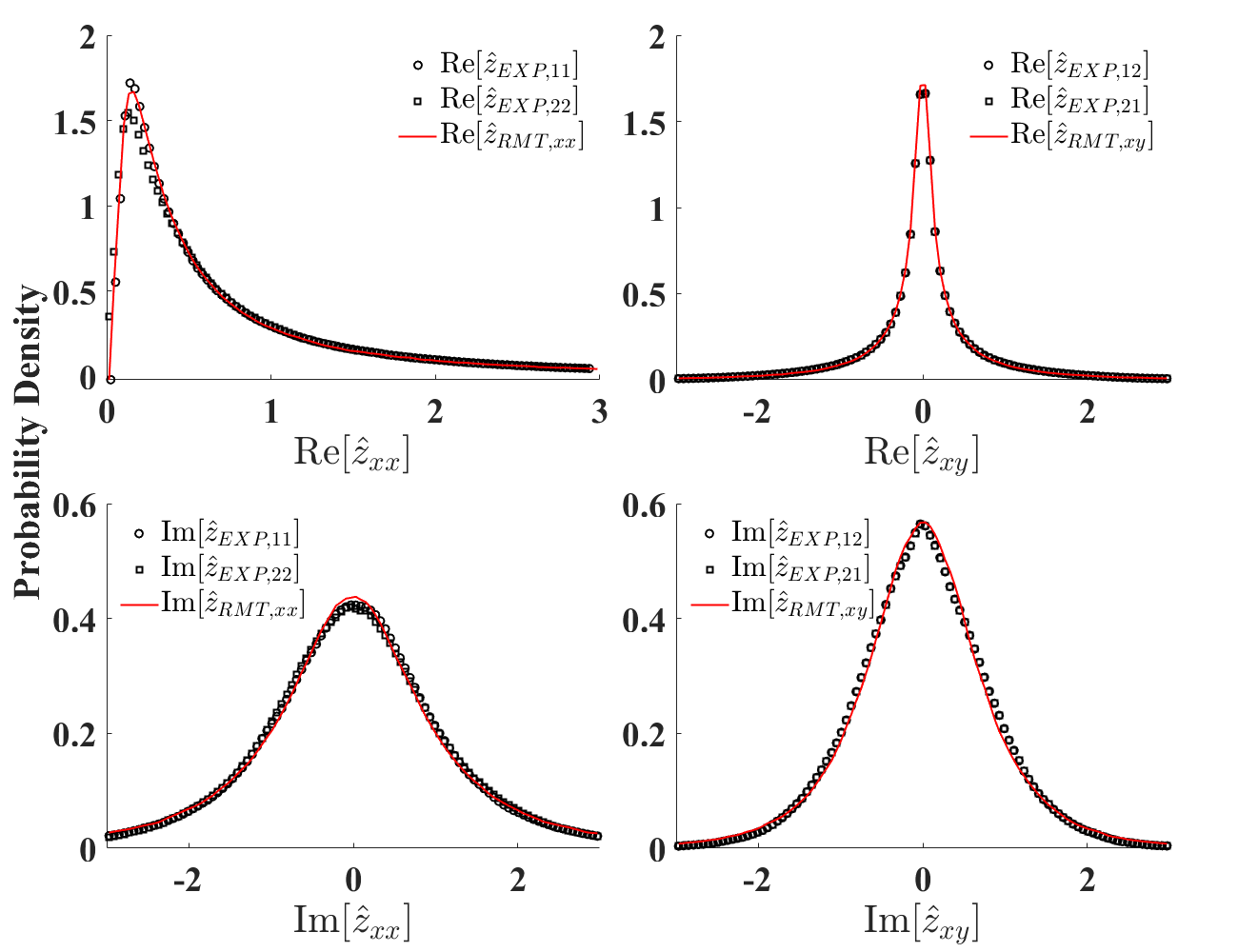}
\caption{\rev{PDFs of the normalized impedance $\hat{z}_{EXP}$ from an experimental tetrahedral graph (black symbols) compared to $\hat{z}_{RMT}$ generated using Eq.~\ref{ZRCM} with $\eta=2.3$ (red line). The experimental data fits well to the numerical distribution, with slight deviations which are understood to be caused by the small size of the graph.}}
\label{RCMStats}
\end{figure}

\rev{We characterize each ensemble measurement with a value of the uniform attenuation $\eta$ using the Random Coupling Model (RCM) \cite{Hemmady2005,Zheng20051Port,Zheng20052Port,Zheng2006,Hemmady2012,Gradoni2014}. The first step in this process is to normalize the data to have perfect coupling through calculation of the normalized impedance $\hat{z}_{EXP}$ from the measured impedance matrix of the cavity $\hat{Z}_{cav}$ by the formula
\begin{equation}
    \hat{Z}_{cav} = i\text{Im}\left[\langle\hat{Z}_{cav}\rangle \right]+\text{Re}\left[\langle\hat{Z}_{cav}\rangle\right]^{1/2}\hat{z}_{EXP}\text{Re}\left[\langle\hat{Z}_{cav}\rangle\right]^{1/2}. \nonumber 
\end{equation}
Next, we compare the statistics of $\hat{z}_{EXP}$ to the numerically generated distributions of normalized impedance $\hat{z}_{RMT}$ calculated using Eq.~\ref{ZRCM}, where $W_n$ is a variable describing the coupling of mode $n$ to each port, and are taken to be Gaussian random variables. Here $k_n$ are the wavenumbers of the cavity modes, and are the eigenvalues of random matrices taken from a Gaussian orthogonal (unitary) ensemble for $\beta=1$ ($\beta=2$). By matching a statistical distribution of $\hat{z}_{RMT}$ to the distribution of $\hat{z}_{EXP}$, we determine the value of $\eta$ for the experimental ensemble. An example of this process can be seen in Fig.~\ref{RCMStats}, Fig.~4 of Ref.~\cite{Fu2017} or Fig.~10 of Ref.~\cite{Tornike2023}.

\begin{equation}
    \hat{z}_{RMT}(k_0) = -\frac{i}{\pi}\sum_{n}\frac{W_nW_n^T}{\frac{k_0^2-k_n^2}{\Delta}+\frac{i\eta}{4\pi}}\label{ZRCM}
\end{equation}

For uniform loss parameter $\eta<12.6$ we utilized generated $\hat{z}_{RMT}$ with a spacing of $\Delta\eta= 0.06$, while for $\eta>12.6$ the spacing was $\Delta\eta= 0.6$. This is a limiting factor on our precision for determination of $\eta$. There is some added uncertainty in $\eta$ based on the goodness of fit between the $\hat{z}_{EXP}$ and $\hat{z}_{RMT}$ distributions.

Because we are extracting the uniform attenuation of a statistical ensemble, rather than just a single cavity realization, we are in a sense averaging the contributions of multiple $\eta$ values. The distribution of $\hat{z}_{RMT}$ has a sensitivity to loss that scales roughly inversely with $\eta$, so it is more difficult to fit $\hat{z}_{EXP}$ to a single value of loss when the distribution is a composite of PDFs of different shapes and variance, which is the case for $\eta\approx5$ and below \cite{Fu2017}.

This can be particularly challenging when working with lower quality ensembles in which the loss between realizations can vary substantially as the global properties such as $L_e$ are changed. Consider that in Fig.~4 of Ref.~\cite{Fu2017}, the authors had to use $\eta=34$ to fit the reflection elements of $\hat{z}_{EXP}$ to a $\hat{z}_{RMT}$, but for the transmission elements they used $\eta=10$. Whereas in Fig.~\ref{RCMStats}, we were able to use $\eta=2.3$ to fit every element of the normalized impedance, for a similar graph. For all the ensembles we included in Table.~\ref{TailTable}, the distribution of $\hat{z}_{RMT}$ matched the data with just a single value of $\eta$, given our spacing stated above.

}

The absorption strength of the systems in this study were in the range of $1.8\leq\eta\leq50$, which extends into what are considered the low- and high-loss regimes. In Fig.~\ref{WS}, it can be seen that the complex $\tau_{WS}$ distributions for systems with larger $\eta$ have the $-3$ power-law tail behavior start at smaller values of $|\text{Re}[\tau_{WS}]|$ and $|\text{Im}[\tau_{WS}]|$. This is similar to a prediction in Ref.~\cite{Lei2021Stats}, that in the ultra-low loss regime ($\eta \ll 1$), an intermediate regime $1\ll M\text{Re}[\tau_{WS}] \ll \frac{1}{\eta}$ exists where the distribution of the real part of the Wigner-Smith time-delay is described by $\mathcal{P}(\text{Re}[\tau_{WS}]) \propto \text{Re}[\tau_{WS}]^{-(M\beta)/2-2}$. Notice that this is the exact same form as predicted for the tail of the distribution of the purely real Wigner-Smith time-delay in the Hermitian case.

From this theory, we can see that there should be a transition from Hermitian-like to non-Hermitian behavior for very small $\eta$ values. This may be explained as modes which have small scattering times experience small amounts of attenuation so the system is almost Hermitian. Those that experience large scattering times experience substantial attenuation, even if $\eta$ is small, and therefore we should expect the non-Hermitian $-3$ power-law behavior.

Experimentally, the lowest loss systems were $\mathcal{D}=1$ graphs measured at very low frequencies. The lowest loss ensemble measured has an estimated absorption of $\eta=1.8$ at $2$ GHz, suggesting that we are bounded by the dielectric loss of our coaxial components. Using RMT, the predicted transition is somewhat visible in Fig.~\ref{RMTWS} when comparing the green curve ($\eta=1.3$) to the blue curve ($\eta=0$) and the dashed red line. The two curves are degenerate for $0.1<\text{Re}[\tau_{WS}]<2$ and are parallel to the dashed red line which characterizes a $-4$ power-law. Note that the finite-loss green curve splits off and becomes parallel to the dashed black line characterizing a $-3$ power-law.

To experimentally observe the predicted transition would likely require a special cryogenic system. That is beyond the scope of this project, and remains an open problem for future research. At the other extreme, we cannot experimentally access the ultra-high loss regime for arbitrarily large $\eta$ values. The tails of the CTD distributions which we are interested in are due to the longest lingering waves. For systems with very large loss, waves that linger for long times are attenuated down into the noise floor of our microwave network analyzer. As a result, we are bounded both above and below on the values of $\eta$ for which we can gather high-quality time-delay statistics.

\subsection{Importance of Frequency Sampling} \label{SupMat_Sampling}

\begin{figure}[ht]
\includegraphics[width=0.48\textwidth]{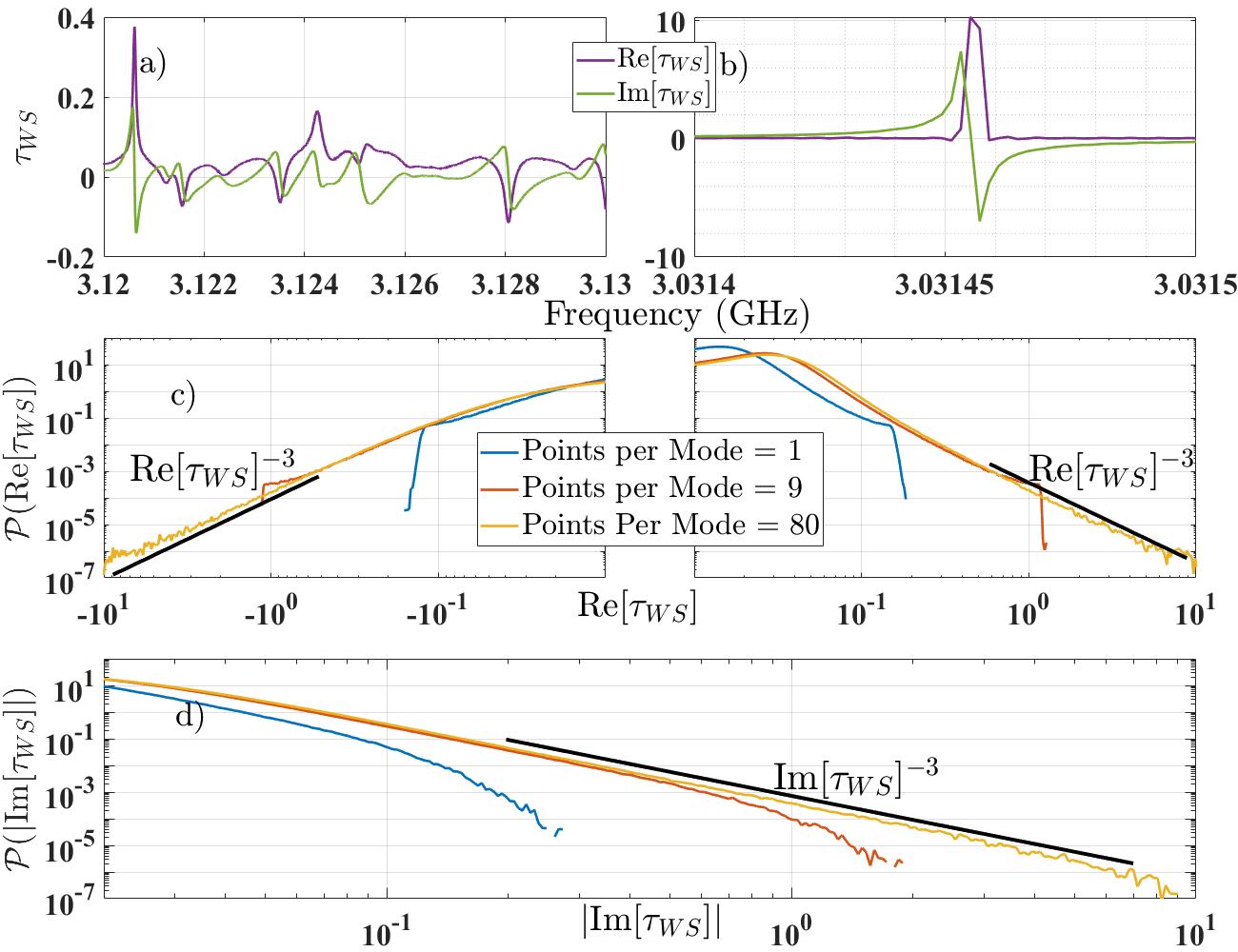}
\caption{Complex Wigner-Smith time-delay data from a $\mathcal{D}=3, M=2$ system. (a) $\tau_{WS}$ as a function of frequency. Both real (in purple) and imaginary (in green) components can be positive and negative, with complicated structure as a function of frequency. (b) Zoom in on a mode with large values for the real and imaginary parts of $\tau_{WS}$, showing how the extreme excursions can be very sharp and narrow, requiring high frequency sampling resolution. (c-d) PDFs of Re$[\tau_{WS}]$ and $\lvert$Im$[\tau_{WS}]$$\rvert$ using three different frequency sampling resolutions. The blue and orange curves show that when the sampling in frequency is not fine enough, the extreme time-delays are not represented in the distribution. Black reference line characterizes asymptotic behavior as power-law tail $\tau_{WS}^{-3}$ for the distribution with high sampling resolution.}
\label{sampling}
\end{figure}

In Fig.~\ref{sampling}(a) we show an example of the complex $\tau_{WS}$ as a function of frequency for one realisation of a three-dimensional microwave cavity. In this frequency range ($3.12-3.13$ GHz), the value of the Wigner-Smith time-delay is small so that the complicated structure of all the modes is visible. The focus of this paper, however, is on the statistics of the tails of the time-delay distributions which correspond to the anomalously long time scattering events. An example of such an event is shown in Fig.~\ref{sampling}(b). This extreme event (note the time-delay scale difference between Figs.~\ref{sampling}(a) and (b)) takes place over a very small frequency range. Therefore to properly measure these events, it is necessary to measure with high resolution in frequency. This can also be seen from Eqs.~\ref{DefWS}-\ref{DefdT}, which have time-delay as an energy (or equivalently, frequency) derivative of the scattering matrix. Because we are taking a frequency derivative, higher resolution data is needed for time-delay determination, as compared to simply resolving the modes of the system directly from the frequency-dependent $S$-matrix. The curve in Fig.~\ref{sampling}(b) is not smooth, showing that this data was undersampled.

In Fig.~\ref{sampling}(c-d) we show what the time-delay distributions look like when the frequency resolution is not sufficiently high. The blue curve corresponds to the distribution calculated from data which only has 1 frequency point per mean mode spacing $\Delta$, the orange curve corresponds to 9 points per mode spacing, and the yellow curve corresponds to 80 points per mode spacing. This approximate order of magnitude increase makes clear how important the frequency sampling is to seeing the time-delay distribution tails. It is important to note that $\Delta$ is not the only parameter which determines the density of sampling points. The degree of absorption $\eta$ and the number of scattering channels $M$ both affect how wide the modes are.

\section{Transmission time-delay difference} \label{SupMat_deltaT}

The transmission time-delay difference is a new quantity introduced for the first time in this paper. 

The most basic way to define the transmission difference is with Eq.~\ref{DefdT}, which is the difference of the transmission time of one path and its time reversed path. Based on the theory presented in Refs. \cite{Osman2020,Lei2022} we can write the complex transmission time-delay difference as a sum over complex transmission zeros $t_n$ for each mode of the closed system $n$:  

\begin{equation}
\begin{aligned}
        \text{Re}[\tau_{\delta T}] = \sum_{n=1}^{N-M}[\frac{\text{Im}[t_n]-\eta}{(E-\text{Re}[t_n])^2+(\text{Im}[t_n]-\eta)^2}\\+\frac{\text{Im}[t_n]+\eta}{(E-\text{Re}[t_n])^2+(\text{Im}[t_n]+\eta)^2}] \label{realtdiff}
\end{aligned}
\end{equation}
\begin{equation}
\begin{aligned}
        \text{Im}[\tau_{\delta T}] = - \sum_{n=1}^{N-M}[\frac{E-\text{Re}[t_n]}{(E-\text{Re}[t_n])^2+(\text{Im}[t_n]-\eta)^2}\\-\frac{E-\text{Re}[t_n]}{(E-\text{Re}[t_n])^2+(\text{Im}[t_n]+\eta)^2}] \label{imagtdiff}
\end{aligned}
\end{equation}

where $\eta$ is the uniform attenuation, $E$ is the energy, $N$ is the number of modes of the closed system, and $M$ is the number of scattering channels.

One thing that is immediately seen in Fig.~\ref{deltaT} is that the peak of $\mathcal{P}(\text{Re}[\tau_{\delta T}])$ is moved away from zero for NRC systems. Whether the peak is moved to a positive or negative values is arbitrary and comes down to which transmission time-delay is subtracted from the other. There is no theory for this qualitatively different behavior, and without more investigation it can't be said whether or not this is a universal feature; the shifted peak could depend on the structure of the graph, for example.

For a perfectly reciprocal system the PDF of the transmission time-delay difference $\tau_{\delta T}$ is identically a delta function at 0, for both the real and imaginary components. Random Matrix Theory numerics and circuit simulation results show this trivial result, but experimental data does not show this behavior as there are always small differences between measured transmission parameters (i.e. complex $S_{21} \neq S_{12}$ in detail, even for nominally reciprocal systems). Such differences could be caused by imperfect calibration, a bias in the measurement device, small magnetic inclusions in the system, etc and are impossible to fully eliminate. The relative differences between $S_{21}$ and $S_{12}$ are found to be enhanced when the transmission amplitudes are vanishingly small. The transmission zeros cause large spikes in both real and imaginary transmission time-delay difference, which can be seen by inspecting the logarithm in Eq.~\ref{DefdT}. If the transmission amplitudes approach zero but one is slightly smaller due to experimental fluctuations, the logarithm can become arbitrarily large, leading to a finite imaginary transmission time-delay difference. The real part of the transmission time-delay difference comes from the phase difference of the transmission parameters, which becomes indeterminate when their magnitudes go to zero. In the case of systems with broken-reciprocity, the transmission time-delay differences are well-defined and non-zero for the vast majority of the data as it is highly unlikely for non-reciprocal transmission amplitudes to both be zero at the same frequency.

\rev{So far we have only discussed the most fundamental transmission time delay difference, but for an $M$ channel system there are $\frac{{M \choose 2}\left({M \choose 2}-1\right)}{2}$ unique transmission time-delay differences that one can define, of five different types (1) between the same two ports (e.g. $\tau_{12}-\tau_{21}$) as in Eq.~\ref{DefdT}, (2) transmission to the same port (e.g. $\tau_{12}-\tau_{13}$), (3) transmission from the same port (e.g. $\tau_{21}-\tau_{31}$), (4) a travelling transmission (e.g. $\tau_{21}-\tau_{32}$), (5) and transmission across completely different ports (e.g. $\tau_{21}-\tau_{43}$). In Fig.~\ref{deltaT} we show only the first kind of transmission time-delay difference, but types (1-4) all have the same $-3$ power-law tails. It should be noted that types (2-4) show $-3$ power-law tails even for \textit{reciprocal} systems. The fifth kind of transmission time delay difference requires $M\geq 4$, which we did not experimentally measure for this study.}

\section{Additional Experimental Complex Time-Delay Statistics Results} \label{SupMat_Individual}

In this section we provide the distributions of the measured complex reflection and transmission time-delays, all of which also show the same $-3$ power-law tail behavior, independent of $\mathcal{D},M,\beta,\eta$. This is the first experimental demonstration of this superuniversal tail, and we are not aware of any theory which predicts or explains these observations. \rev{We also show the distribution of the complex reflection time-delay difference.}

\begin{figure}[bht]
\includegraphics[width=0.48\textwidth]{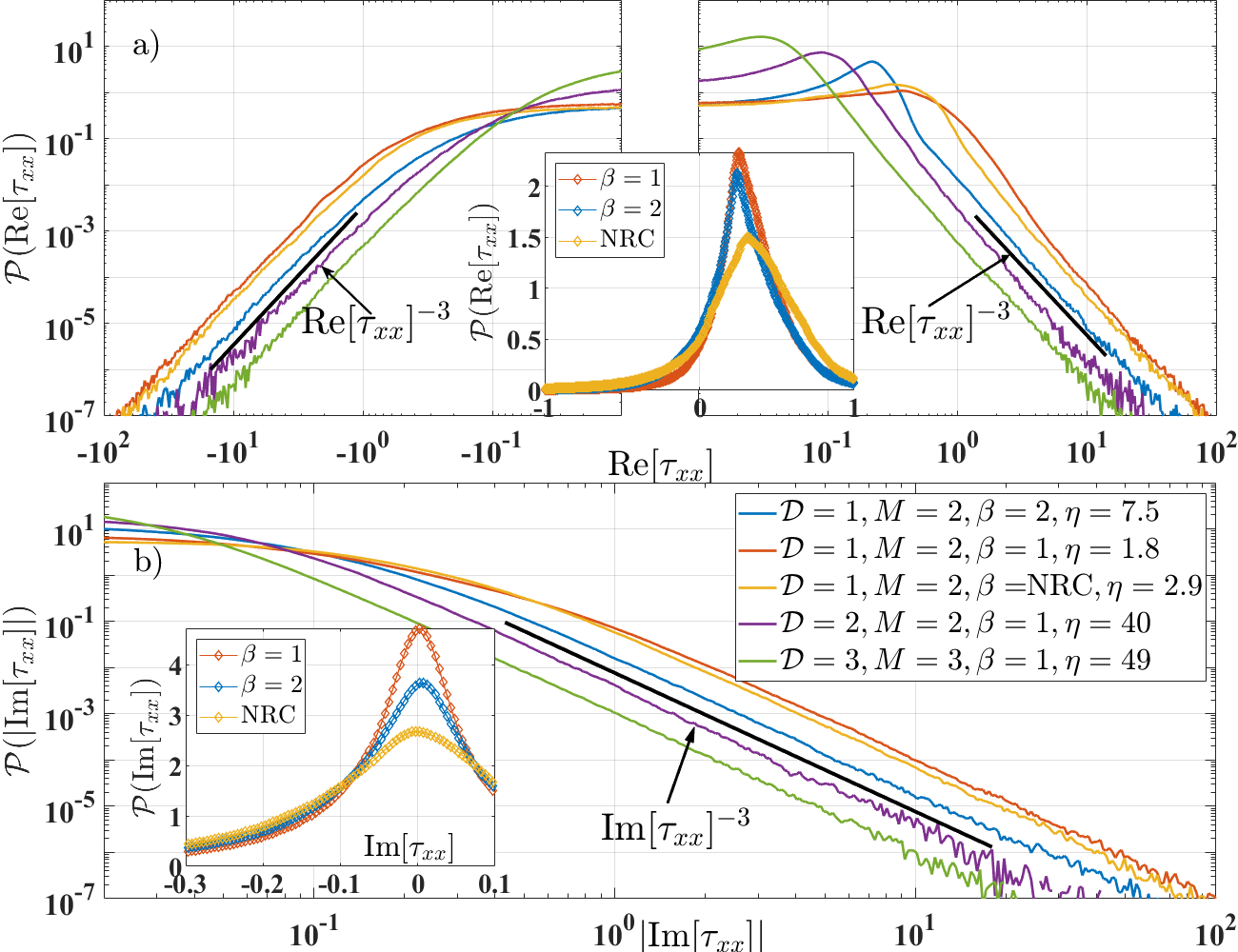}
\caption{PDFs of (a) Re$[\tau_{xx}]$ and (b) $\lvert$Im$[\tau_{xx}]$$\rvert$ reflection CTD from select ensembles of chaotic scattering systems with different values for the four parameters $\mathcal{D},M,\beta,\eta$. Insets and reference lines serve the same purpose as in Fig.~\ref{WS}. }
\label{Rxx}
\end{figure}

\begin{figure}[ht]
\includegraphics[width=0.48\textwidth]{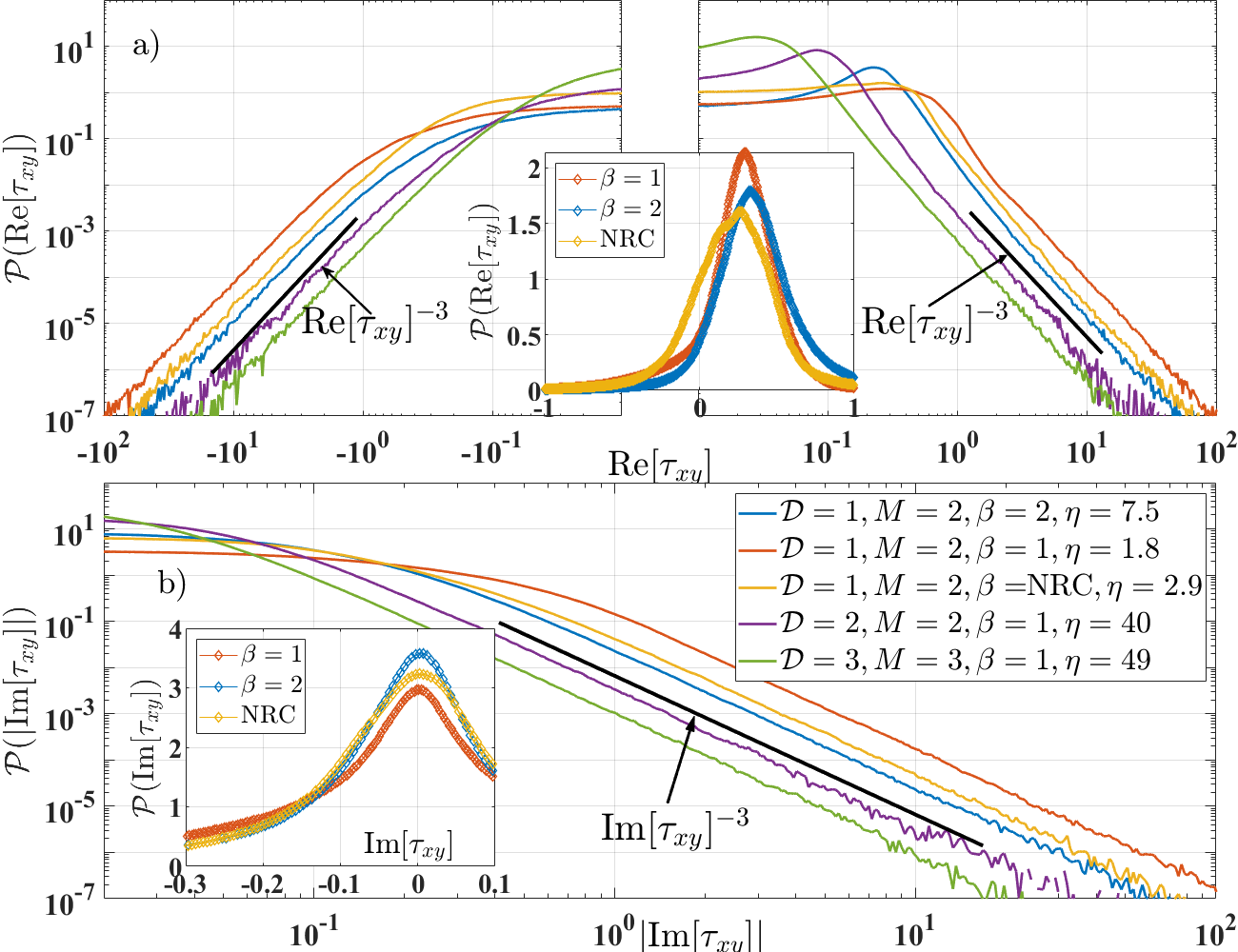}
\caption{PDFs of (a) Re$[\tau_{xy}]$ and (b) $\lvert$Im$[\tau_{xy}]$$\rvert$ transmission CTD from select ensembles of chaotic scattering systems with different values for the four parameters $\mathcal{D},M,\beta,\eta$. Insets and reference lines serve the same purpose as in Fig.~\ref{WS}. }
\label{Txy}
\end{figure}

Fig.~\ref{Rxx} shows the distributions of the complex reflection time-delays $\tau_{xx}$ for five systems with different parameter values, including $M=2$ and $M=3$. As discussed in the main text, the distribution of the tails of the reflection time-delay can be explained by equating it to the Wigner-Smith time-delay of a $M=1$ system, but the distributions of the transmission time-delays $\tau_{xy}$ ($x\neq y$), shown in Fig.~\ref{Txy}, require a different analysis to be understood.


\begin{figure}[htb]
\includegraphics[width=0.48\textwidth]{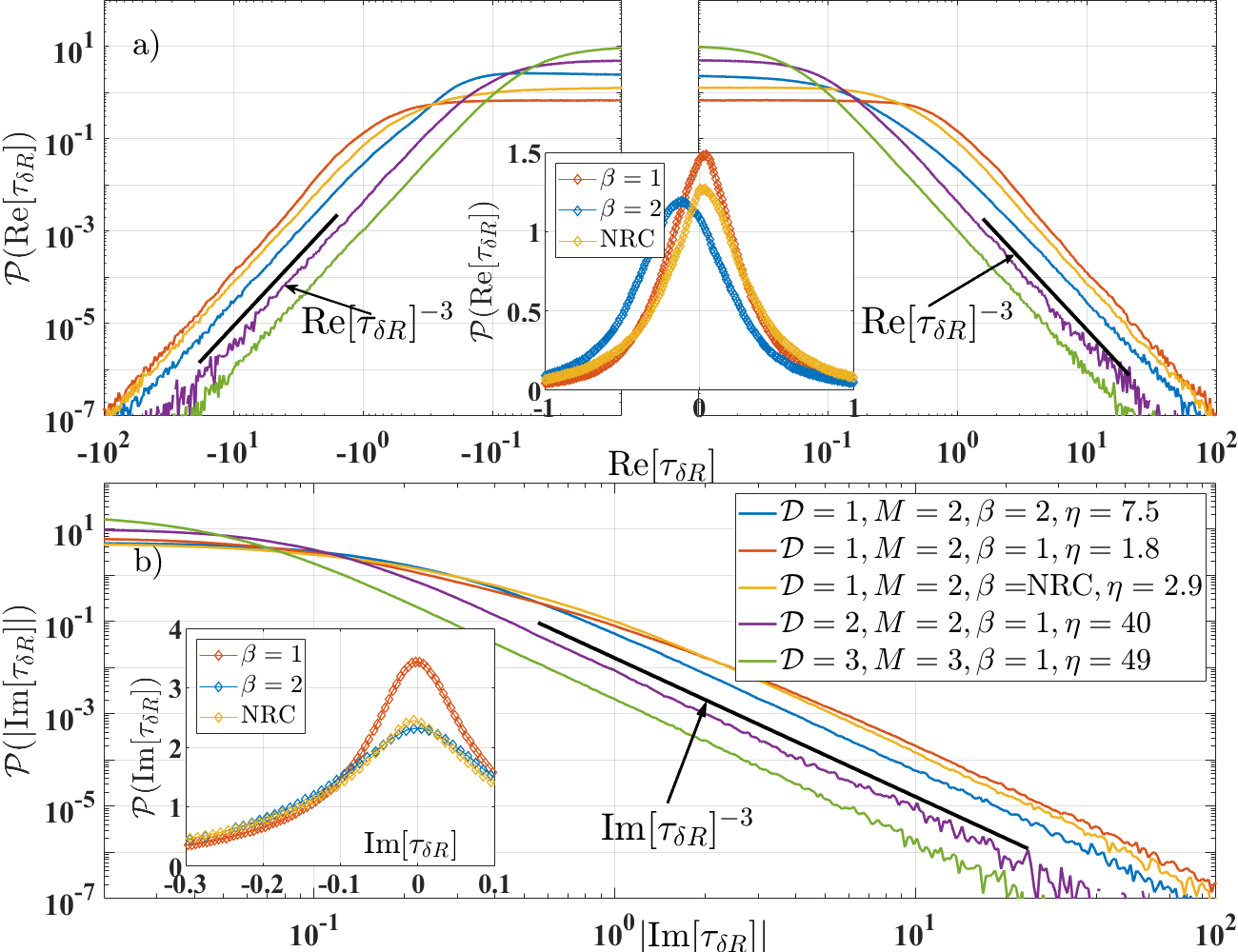}
\caption{PDFs of (a) Re$[\tau_{\delta R}]$ and (b) $\lvert$Im$[\tau_{\delta R}]$$\rvert$ reflection time-delay difference from ensembles of experimental scattering systems with different values for the four parameters $\mathcal{D},M,\beta,\eta$. Insets and reference lines serve the same purpose as in Fig.~\ref{WS}. }
\label{deltaR}
\end{figure}

\rev{An $M$ channel system has $\frac{M(M-1)}{2}$ reflection time-delay differences. For the $M=3$ systems measured in the experiment it was found that the distributions of each pairwise difference have identical tails with only slight differences in the shape of the PDF at small values of $\tau_{\delta R}$. For that reason we only present the distribution of one difference for the $M=3$ system in Fig.~\ref{deltaR}, which is the green line. }

\section{Random Matrix Theory Numerics with Loss and Gain} \label{SupMat_RMT}

In this section, we demonstrate that numerical data from Random Matrix Theory numerical simulations displays the same superuniversal $-3$ power-law tails for all CTD distributions. The fact that scattering matrices created from random matrix ensembles show the same power-law behavior demonstrates that the power-law tails are generic. We use ensembles of $300$ Gaussian random matrices that are of dimension $10^5\cross10^5$. The process of creating the random matrices and using the eigenvalues to generate frequency dependent $S$-matrix data is detailed in Ref.~\cite{Lee2013}, Appendix A of Ref.~\cite{Hemmady2012}, and Appendix A of Ref.~\cite{LeeThesis}. Accounting for the sampling requirement described in Section \ref{SupMat_Sampling}, we calculate the $S$-matrices very finely in the frequency domain, with 336 points per mean mode spacing.

\begin{figure}[ht]
\includegraphics[width=0.48\textwidth]{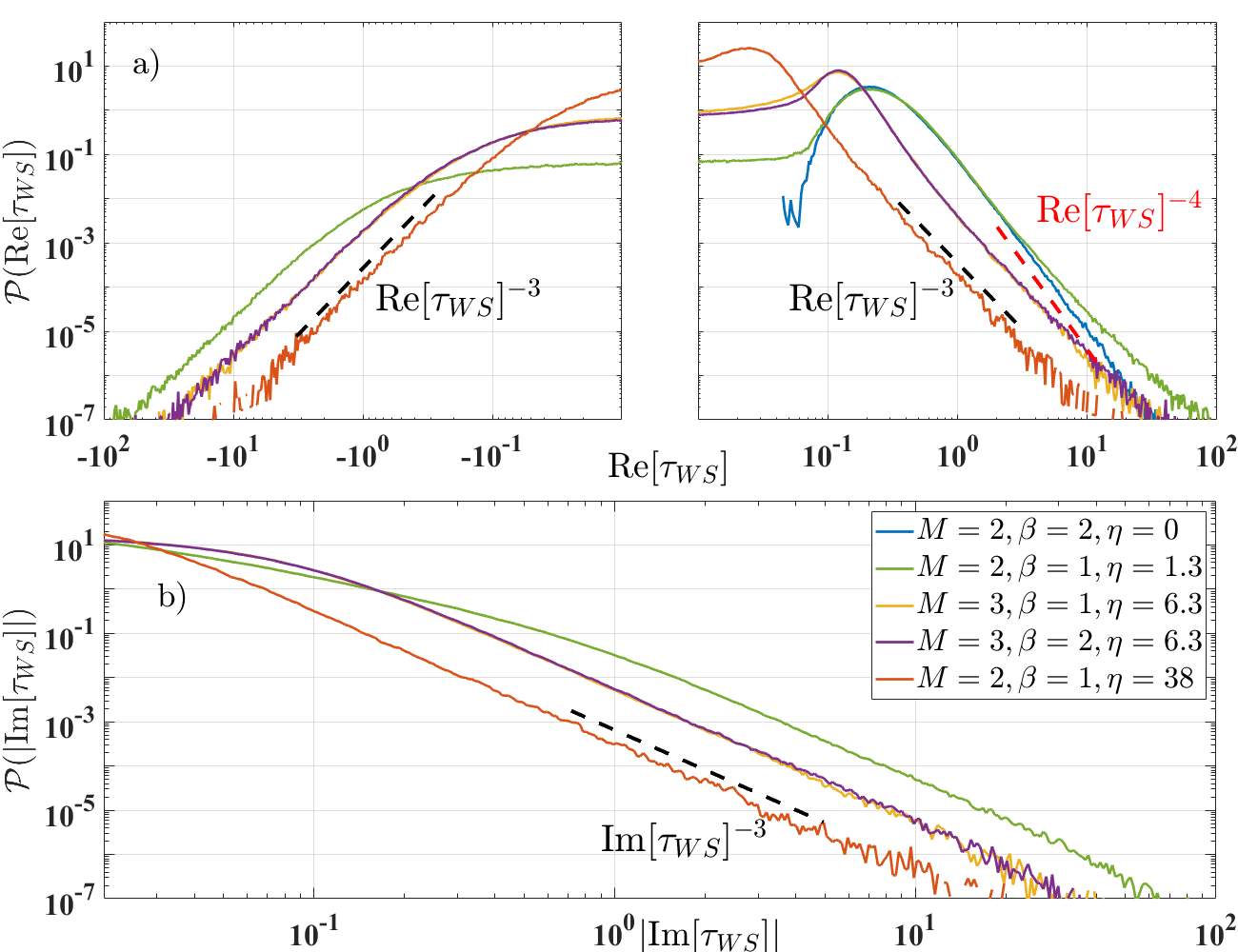}
\caption{PDFs of (a) Re$[\tau_{WS}]$ and (b) $\lvert$Im$[\tau_{WS}]$$\rvert$ Wigner-Smith CTD from RMT numerics with different values for $M,\beta,\eta$. Black reference line characterizes asymptotic behavior as power-law tail $\tau_{WS}^{-3}$ for distributions with $\eta>0$. Dashed red line characterizes zero loss asymptotic behavior as $\tau_{WS}^{-\frac{M\beta}{2}-2}$, where $-\frac{M\beta}{2}-2=-4$ for the blue line.}
\label{RMTWS}
\end{figure}

In Fig.~\ref{RMTWS} we show Wigner-Smith time-delay statistics based on RMT numerics, and the blue curve corresponds to a $M=2,\beta=2,\eta=0$ system, and has a $-4$ power-law tail as predicted for Hermitian scattering systems. The Wigner-Smith time-delay for this system is purely real and positive. The green curve corresponds to a $M=2,\beta=2,\eta=1.3$ system, and is very similar to the lossless case for a range of $\text{Re}[\tau_{WS}]$ before splitting off to have a $-3$ power-law tail. This plot displays the transition region in the distribution discussed in Sec.~\ref{SupMat_Systems}. We also see what was noted in the main text, that the distributions from systems with larger $\eta$ have the power-law tail onset at smaller values of time-delay. The yellow and purple lines both have $M=3,\eta=6.3$, but one corresponds to a reciprocal system and the other does not. The fact that they are essentially on top of each other throughout the entire range plotted shows that $\beta$ has no relevance to the CTD statistics of non-Hermitian systems, whereas it is explicitly dependent on this parameter for Hermitian ones.

With our microwave experiments we are limited to studying only passive lossy systems, but with RMT it is as simple as taking $\eta \rightarrow -\eta$ to simulate a system with gain instead of loss. We show in Supp. Mat. \cite{SuppMatt} Sec.~\ref{SupMat_RMT} that at least in the case of RMT numerics the distributions of the CTDs maintain the $-3$ power-law tails even in a system with gain. For the Wigner-Smith time-delay, it is not just the tail that remains the same upon going from loss to gain; in a system with no lumped loss the Wigner-Smith time-delay with uniform absorption $\eta$ is the complex conjugate of the time-delay with gain $-\eta$ ($\tau_{WS,\eta}^{\dag}=\tau_{WS,-\eta}$). This can be seen immediately when writing the Wigner-Smith time-delay in the form of a sum over modes of Lorentzians terms involving the poles and zeros of the $S$-matrix \cite{Lei2021WS}. Because we haven't done any experimental work on systems with gain, we make no claims on the validity or implications of the distributions with negative $\eta$ values.

\begin{figure}[ht]
\includegraphics[width=0.48\textwidth]{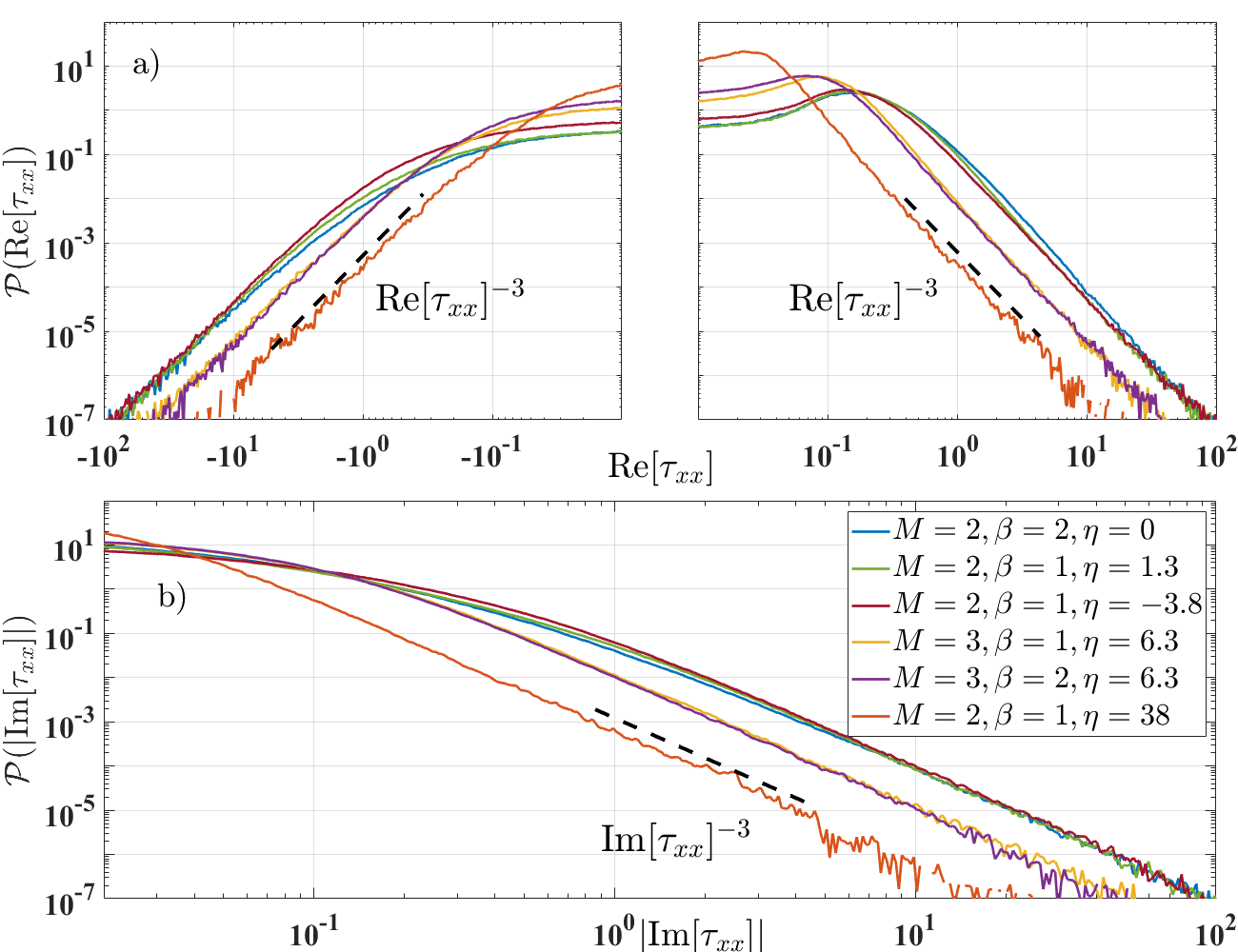}
\caption{PDFs of (a) Re$[\tau_{xx}]$ and (b) $\lvert$Im$[\tau_{xx}]$$\rvert$ reflection CTD from RMT numerics with different values for $M,\beta,\eta$. Black reference line characterizes asymptotic behavior as power-law tail $\tau_{xx}^{-3}$, common to all distributions.}
\label{RMTR11}
\end{figure}

\begin{figure}[ht]
\includegraphics[width=0.48\textwidth]{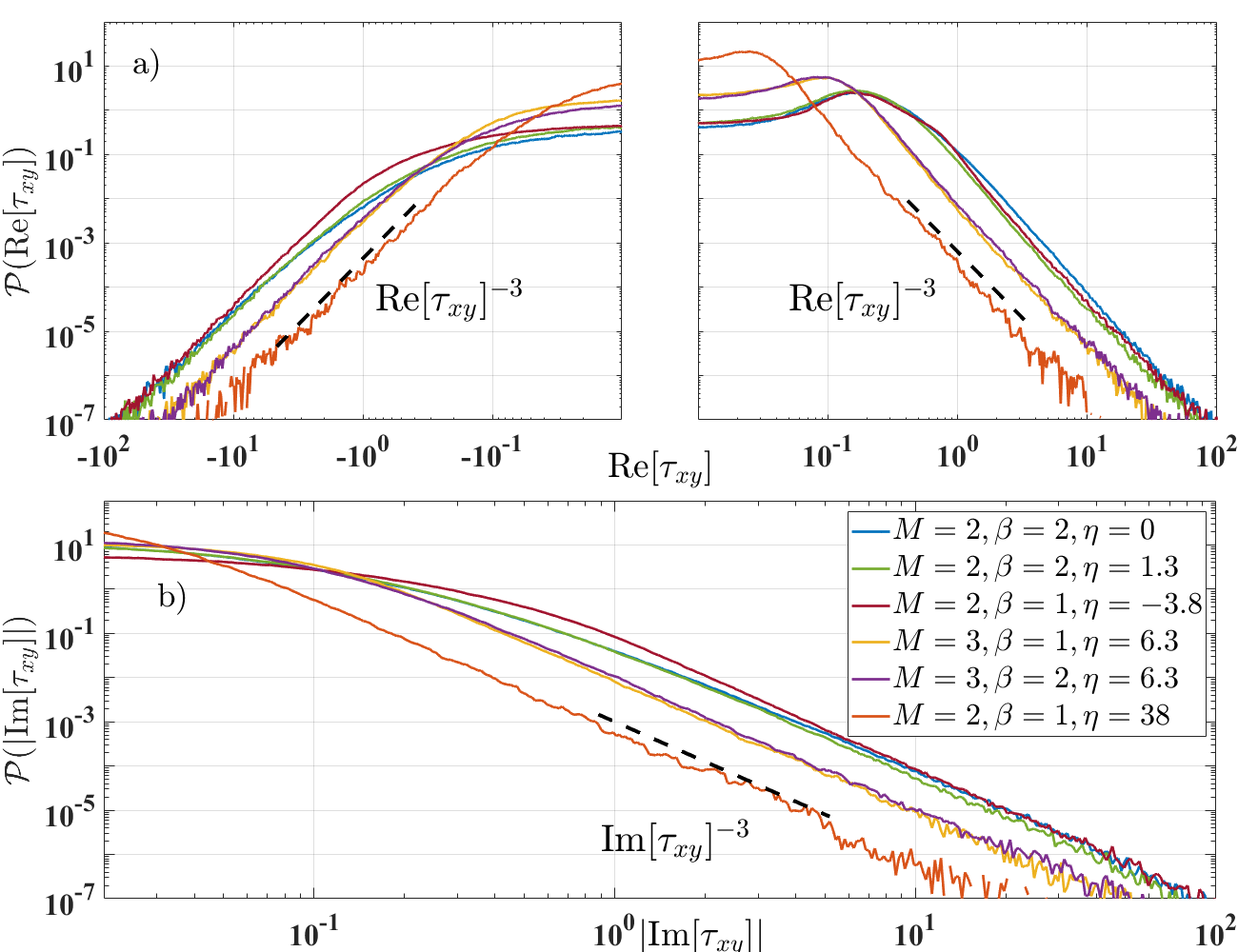}
\caption{PDFs of (a) Re$[\tau_{xy}]$ and (b) $\lvert$Im$[\tau_{xy}]$$\rvert$ transmission CTD from RMT numerics with different values for $M,\beta,\eta$. Black reference line characterizes asymptotic behavior as power-law tail $\tau_{xy}^{-3}$, common to all distributions.}
\label{RMTT12}
\end{figure}

Fig.~\ref{RMTR11} shows the complex reflection time-delay distributions calculated through RMT numerics. Even though the blue curve is for a lossless system ($\eta=0$), it still has a $-3$ power-law tail for both its real and imaginary components of $\tau_{xx}$. This follows from the previously stated idea that the reflection time-delay of an $M$ channel system is the same as the Wigner-Smith time-delay of that same system with 1 scattering channel and $M-1$ absorbing channels. Similarly, Fig.~\ref{RMTT12} shows the distributions for the complex transmission time-delay $\tau_{xy}$ ($x\neq y$), all of which have $-3$ power-law tails. Even for a Hermitian system with a unitary $S$-matrix, the reflection and transmission sub-matrices are generically subunitary and therefore the reflection and transmission time-delays are necessarily complex and display non-Hermitian statistics.

\begin{figure}[ht]
\includegraphics[width=0.48\textwidth]{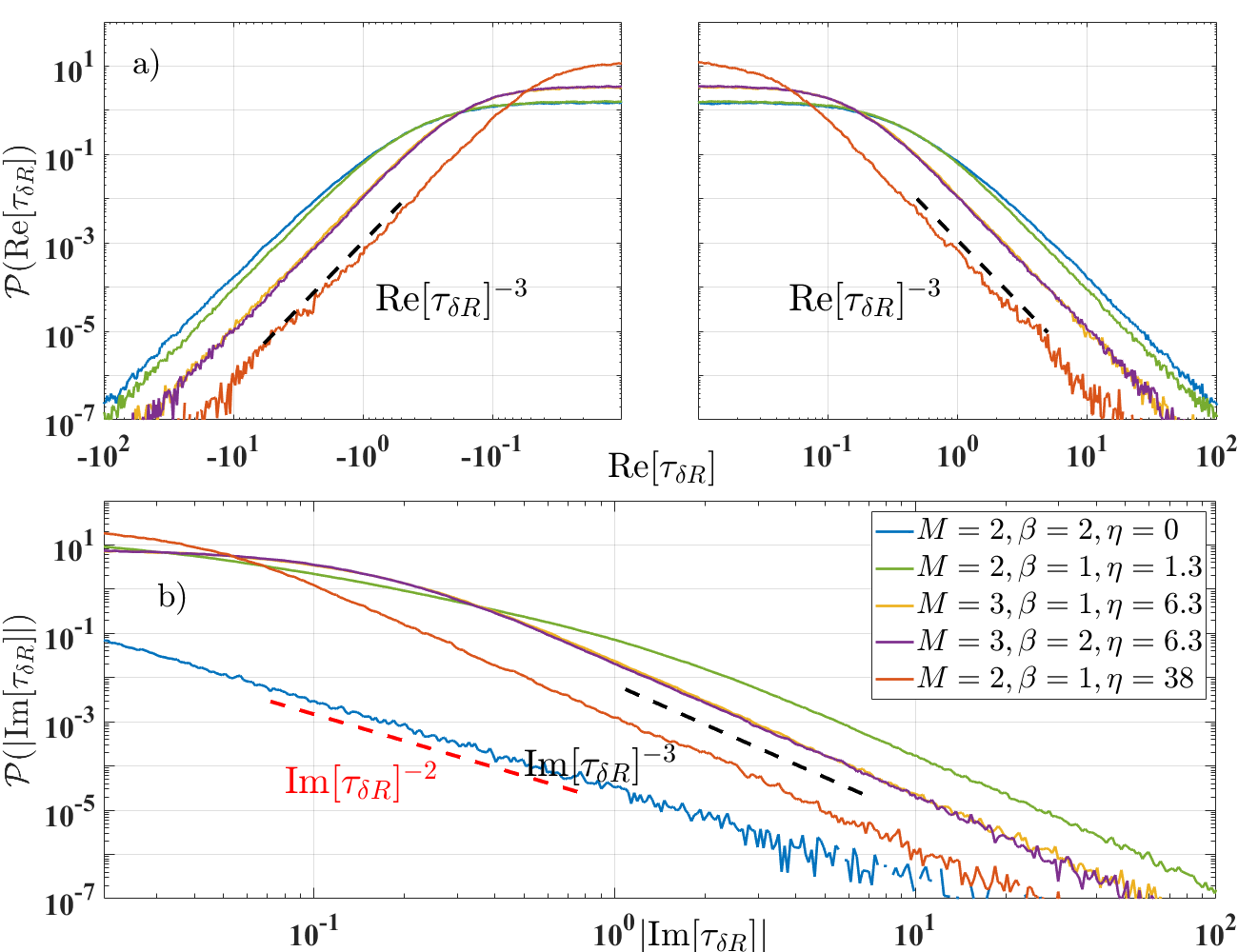}
\caption{PDFs of (a) Re$[\tau_{\delta R}]$ and (b) $\lvert$Im$[\tau_{\delta R}]$$\rvert$ reflection time-delay differences from RMT numerics with different values for $M,\beta,\eta$. Black reference line characterizes asymptotic behavior as power-law tail $\tau_{\delta R}^{-3}$, common to all distributions except $\mathcal{P}_{\eta=0}$(Im$[\tau_{\delta R}]$) (lossless case $\eta=0$) which has a $-2$ power-law characterized by the red reference line.}
\label{RMTRDiff}
\end{figure}

\begin{figure}[ht]
\includegraphics[width=0.48\textwidth]{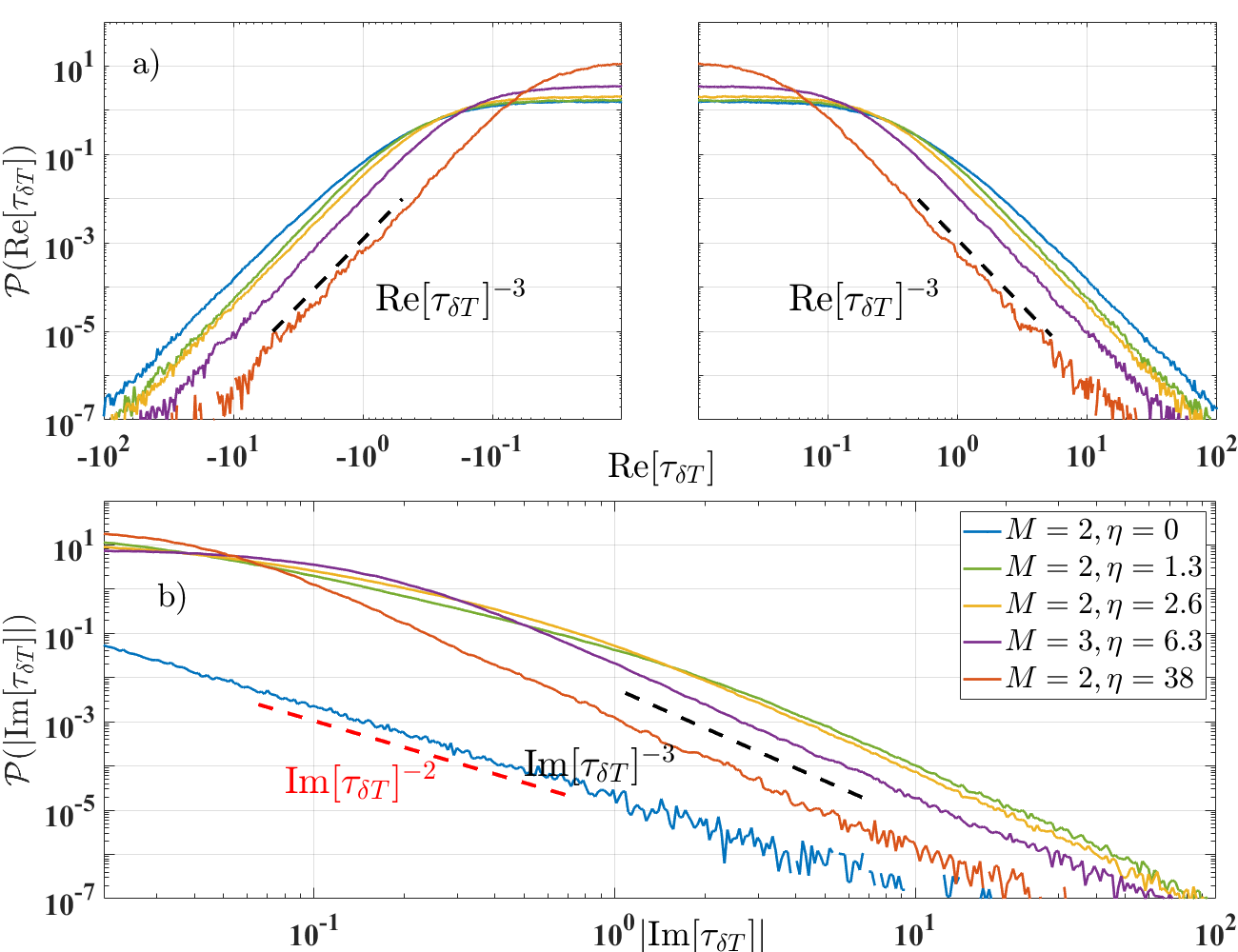}
\caption{PDFs of (a) Re$[\tau_{\delta T}]$ and (b) $\lvert$Im$[\tau_{\delta T}]$$\rvert$ transmission time-delay differences from RMT numerics with different values for $M$ and $\eta$. Black reference line characterizes asymptotic behavior as power-law tail $\tau_{\delta R}^{-3}$, common to all distributions except $\mathcal{P}_{\eta=0}($Im$[\tau_{\delta T}]$) (lossless case $\eta=0$) which has a $-2$ power-law characterized by the red reference line.}
\label{RMTTDiff}
\end{figure}

Figs.~\ref{RMTRDiff} and \ref{RMTTDiff} show the distributions of the complex reflection and transmission time-delay differences based on RMT numerics. It is interesting to note the $-2$ power-law for the imaginary component of the reflection and transmission time-delay differences for $\eta=0$ (unitary scattering system) in the broken reciprocity case. For a lossless, and reciprocal $M=2$ system, it is necessary that $S_{11}=S_{22}$ up to a phase, so the reflection time-delay difference is purely real. So in this case we only see an imaginary component of the time-delay differences for $\beta=2$ at zero loss. When the same calculation is done using $M=3$, however, the distributions of the imaginary time differences show $-3$ tails for both $\beta=1$ and $\beta=2$. This is another clear case of how the statistics for a Hermitian system can have strong parameter dependence, but they become simpler in the non-Hermitian case. 

\section{Limits of Superuniversal Statistics} \label{SupMat_Lim}

In this section, we explore two ways to create deviations from the superuniversal tail behavior that we have discovered. The two effects arise from adding strong lumped loss to the system, and creating very weak coupling between the system and the scattering channels.

\subsection{Lumped Loss and the Disruption of Superuniversal Statistics} \label{SupMat_Lumped}

A question that might arise is how the CTD statistics are affected by lumped loss as opposed to uniform loss? The answer is rather nuanced, and depends on the relative strength of the two types of absorption. If the lumped attenuation is strong, it can disrupt the expected $-3$ tails for the distributions of the complex time-delays. This is because strong lumped attenuation can effectively deform the cavity and bias the wave propagation. The power-law tails depend on the system being ergodic, which won't be the case with large lumped attenuation. However, if the system has some uniform attenuation in addition to the lumped loss, then the lumped attenuation won't be as severe a perturbation, and the statistics reported in this paper can survive. 

We attempted to demonstrate this effect experimentally using a graph with an attenuator attached to one internal node. The attenuator is followed by a short circuit, which reflects the wave back through the attenuator and into the graph. A graph with uniform attenuation $\eta=2.7$ thus had a short to ground through either a 4, 8, or 12 dB attenuator attached to one node. In all cases the $-3$ power-law tails survived in all all types of CTD. When the same system was simulated using a circuit model in CST Studio Suite to have uniform attenuation $\eta=0$, the $-3$ power-law tails were not reproduced, even with just a 4 dB load. Simulations with $\eta=2.7$ did produce the $-3$ power-law tails even with an 8 dB lumped attenuator; the threshold of uniform versus lumped attenuation required to maintain the -3 power-law tails is still an open question.

\subsection{Dependence of Statistics on Coupling} \label{SupMat_Coup}

\begin{figure}[h]
\includegraphics[width=0.48\textwidth]{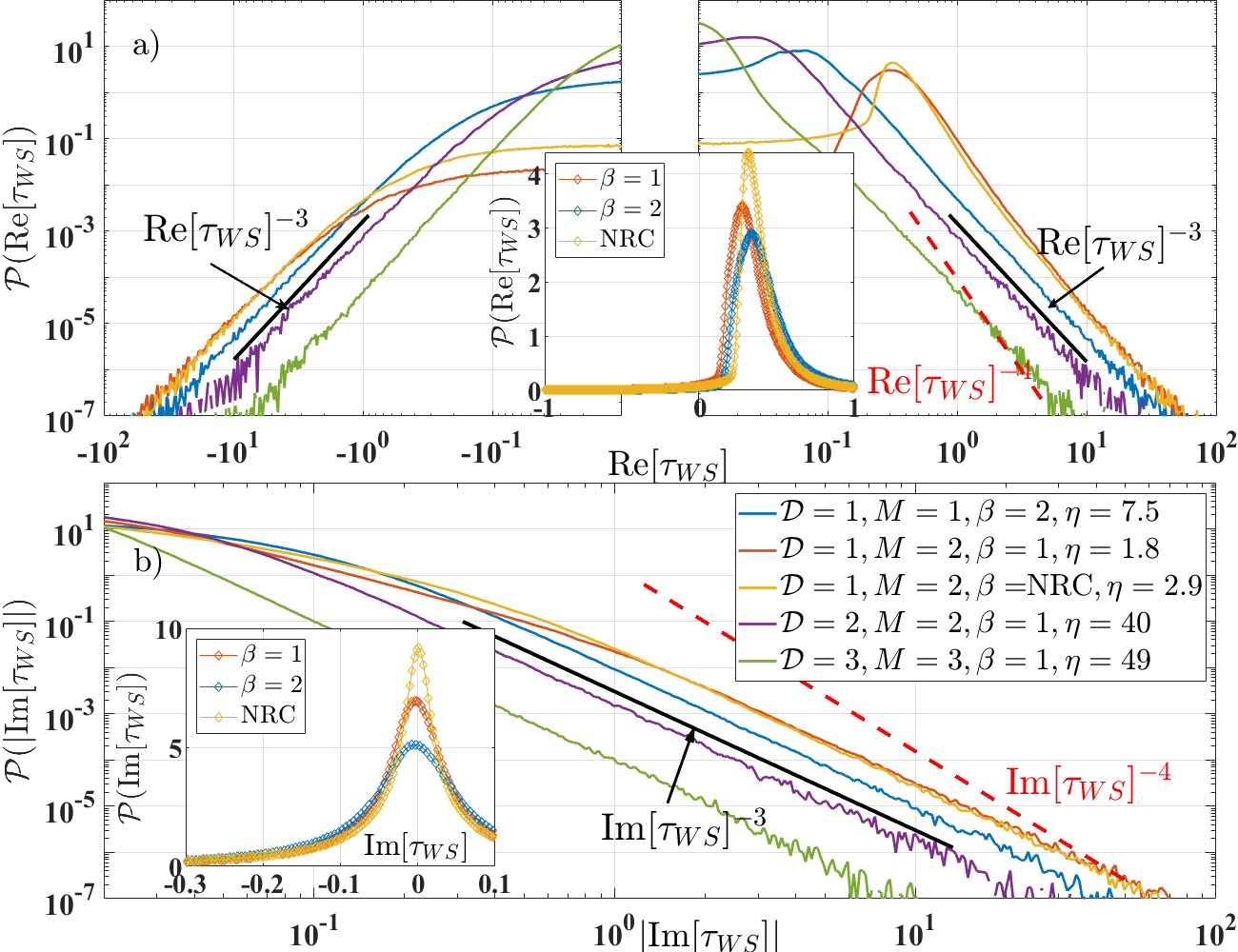}
\caption{PDFs of (a) Re$[\tau_{WS}]$ and (b) $\lvert$Im$[\tau_{WS}]$$\rvert$ Wigner-Smith CTD from select measured $S$-matrix ensembles of chaotic scattering systems with different values for the four parameters $\mathcal{D},M,\beta,\eta$, and imperfect coupling. Insets and reference lines serve the same purpose as in Fig.~\ref{WS}.}
\label{WScoup}
\end{figure}

Here we show the distributions of the CTD calculated from $S$-matrix data with imperfect coupling. This is the same data as shown in the main text and in Sup. Mat. Sec.~\ref{SupMat_Individual}, but the time-delays are calculated using the raw $S$-matrix measured by the network analyzers before RCM-normalization. We find that perfect coupling is \textit{not} required to observe the superuniversal $-3$ power-law tails, as can be seen in Figs.~\ref{WScoup} (Wigner-Smith), ~\ref{T12coup} (transmission), and \ref{RDiffcoup} (reflection time-delay difference).

\begin{figure}[h]
\includegraphics[width=0.48\textwidth]{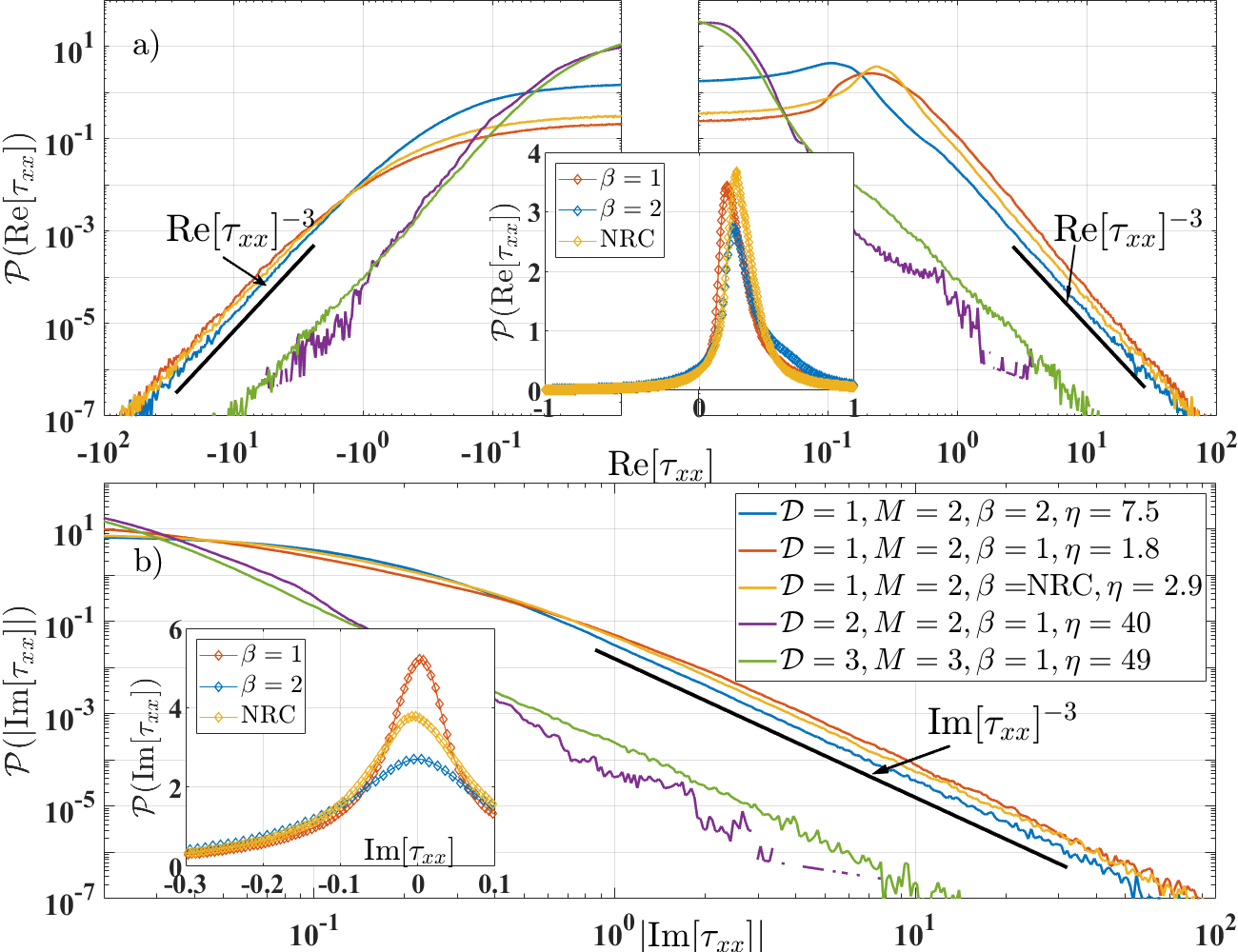}
\caption{PDFs of Re$[\tau_{xx}]$ and (b) $\lvert$Im$[\tau_{xx}]$$\rvert$ port-2 reflection CTD from select ensembles of chaotic scattering systems with different values for the four parameters $\mathcal{D},M,\beta,\eta$, and imperfect coupling. Insets and reference lines serve the same purpose as in Fig.~\ref{WS}.}
\label{Rxxcoup}
\end{figure}

\begin{figure}[h]
\includegraphics[width=0.48\textwidth]{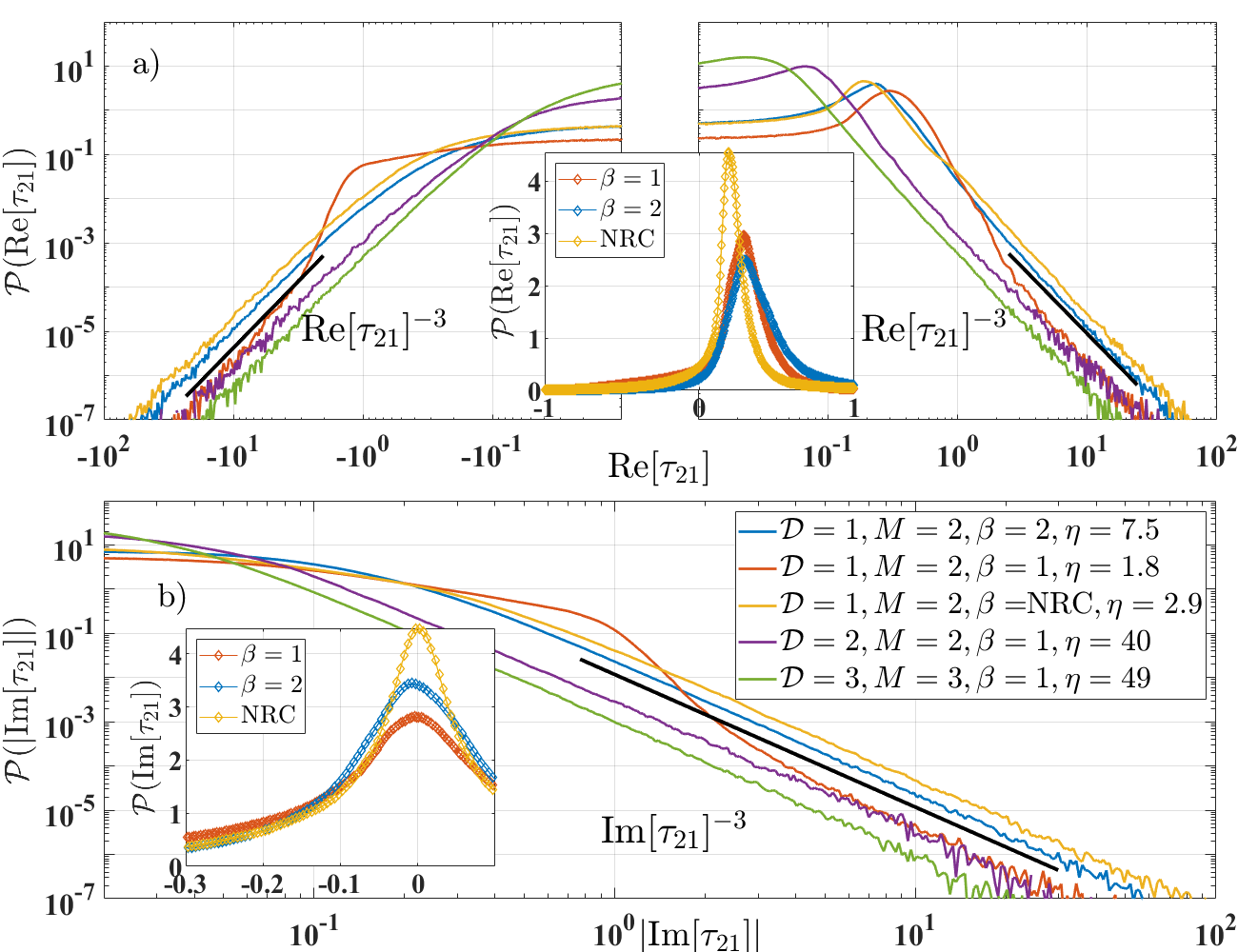}
\caption{PDFs of (a) Re$[\tau_{21}]$ and (b) $\lvert$Im$[\tau_{21}]$$\rvert$ transmission CTD from select ensembles of chaotic scattering systems with different values for the four parameters $\mathcal{D},M,\beta,\eta$, and imperfect coupling. Insets and reference lines serve the same purpose as in Fig.~\ref{WS}.}
\label{T12coup}
\end{figure}

\begin{figure}[h]
\includegraphics[width=0.48\textwidth]{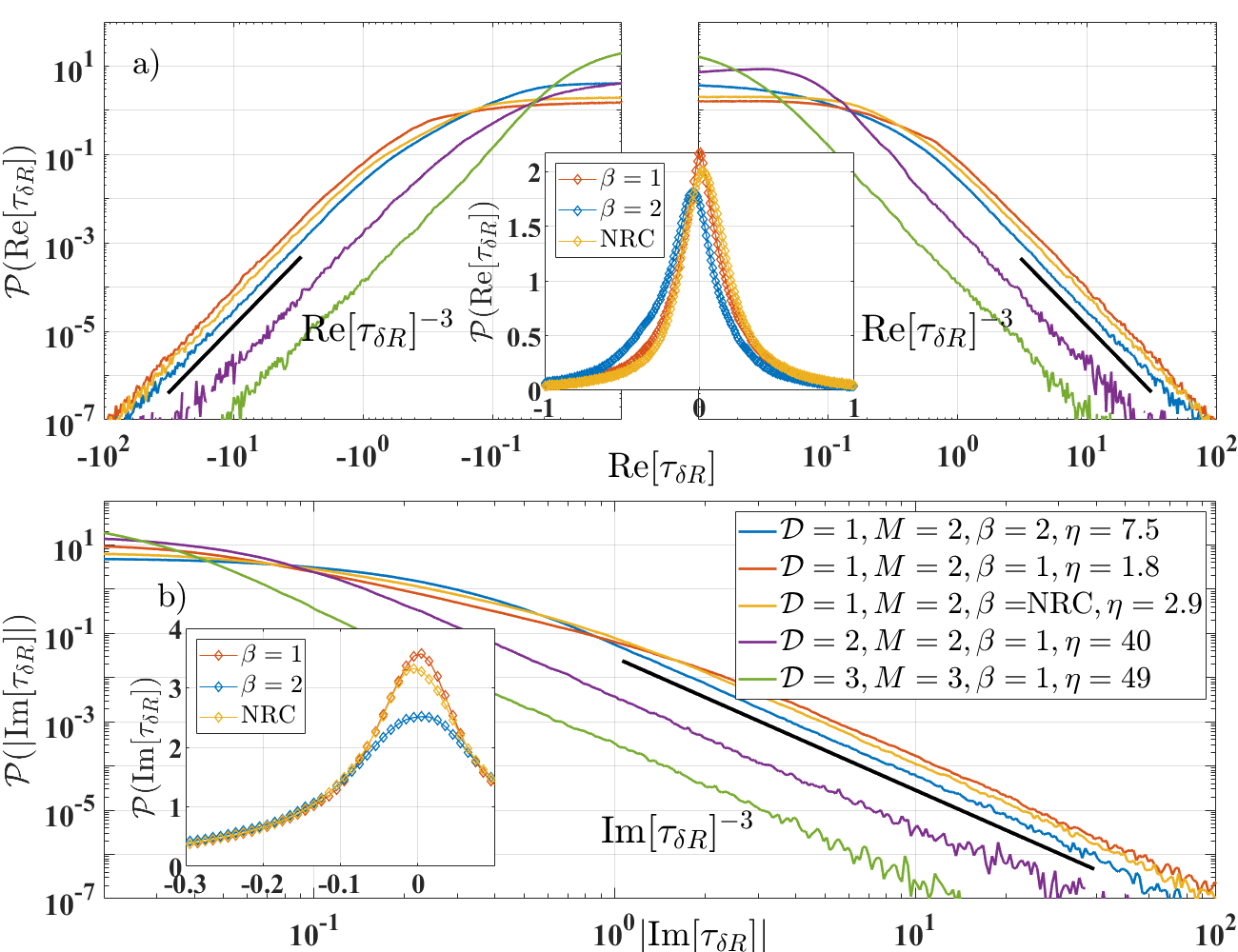}
\caption{PDFs of (a) Re$[\tau_{\delta R}]$ and (b) $\lvert$Im$[\tau_{\delta R}]$$\rvert$ reflection time-delay differences from select measured $S$-matrix ensembles of chaotic scattering systems with different values for the four parameters $\mathcal{D},M,\beta,\eta$, and imperfect coupling. Insets and reference lines serve the same purpose as in Fig.~\ref{WS}.}
\label{RDiffcoup}
\end{figure}

However, Fig.~\ref{Rxxcoup} (reflection time-delay) shows why perfect coupling is important, nonetheless. If the scattering channels are poorly coupled to the cavity, there will be many so called ``direct processes," meaning the waves will immediately be reflected back, and thus not enter the system and suffer a non-zero reflection time-delay. This will cause the reflection time-delay distribution to be very peaked at small time values and not have enough events at large times to show the expected tail behavior. Each channel will have it's own unique degree of coupling to the system, meaning that the reflection time delay distribution at one port might have the superuniversal $-3$ power-law tail while the distribution from a different channel of the same system might not.  It is no great surprise that by purposefully making the coupling at certain channels worse than at others, one can comparatively reduce the likelihood of having RSMs at those channels.  However, it is still interesting to note that even a system with imperfect coupling may still show the $-3$ power-law tails in some if not all of its reflection time-delay distributions, and seemingly will always have the tail in the other CTD quantities.

We do not yet have a quantifiable measure of exactly how good the coupling must be before the $-3$ power-law tail is able to be seen in the reflection time-delay distribution, and it likely depends on the frequency sampling resolution (See Supp. Mat. \cite{SuppMatt} Sec.~\ref{SupMat_Sampling}).

The $\mathcal{D}=1$ graphs have very good coupling even before RCM-normalization because the microwave signals are fed directly into the $M$ nodes using an SMA T-adapter. For the $\mathcal{D}=2$ billiard and $\mathcal{D}=3$ cavity, we have to use antennae which are only well-matched for a small window of the frequency band in which we measure. This is why the green and purple curves are so different from the rest in Fig.~\ref{Rxxcoup}. We chose not to show the distributions of the transmission time-delay difference without perfect coupling because all the non-reciprocal systems in this study were graphs, which, as stated before, have very good coupling even before RCM-normalization.

\section{Extension of Existing Theory} \label{SupMat_Theory}

\rev{We can extend the existing theory for the $-3$ power-law tails of the complex Wigner-Smith time-delay to the complex reflection time-delay through the following argument. The complex reflection time-delay of an $M>1$ system is the same as $\tau_{WS}$ of an $M=1$ system with an additional $M-1$ parasitic channels, increasing the loss in a spatially non-uniform manner. As long as the lumped loss is not much greater than the uniform loss, the results in Supplemental Material \cite{SuppMatt} Sec.~\ref{SupMat_Lumped} suggest that the power-law tails remain. A single hidden channel should not introduce significant loss. Having multiple hidden channels will spread this loss around the system, effectively increasing uniform absorption by way of parasitic channels. This is how uniform absorption is modeled in the Heidelberg formalism \cite{Buttiker1986,Brouwer1996,Kuhl2013}. It then seems reasonable that $\mathcal{P}(\tau_{xx})$ will have the same superuniversal features as $\mathcal{P}(\tau_{WS})$. This can be generalized to any number $m<M$, meaning the CTD of the reflection sub-matrix $R$, which is a diagonal $m\cross m$ block of the $S$-matrix \cite{Faul2024}, should have distributions with $-3$ power-law tails.

Similarly, the argument laid out in the Appendix of \cite{Osman2020} for the $-3$ power-law behavior of $\mathcal{P}[\tau_{\delta R}]$ can be repeated exactly for the transmission time-delay difference $\mathcal{P}[\tau_{\delta T}]$ by substituting Eqs.~\ref{realtdiff}, \ref{imagtdiff} in place of their equation (10). This results in an analysis focused on the transmission zeros $t_n$ rather than the reflection zeros $r_n$. That leaves just the individual complex \textit{transmission} time-delays as lacking a theoretical explanation for the power-law tail. The transmission sub-matrix zeros and scattering matrix poles, which the transmission time-delay depends on, do not have a simple relationship \cite{Huang2022,Huang2022_note}, and we cannot infer the properties of the transmission time-delay from any other CTD quantities.

}

\clearpage
\newpage

\bibliography{StatsCTD.bib}
\nocite{LeeThesis}
\nocite{Huang2022_note}






\end{document}